\documentclass[aps, reprint, superscriptaddress, prb, 10pt]{revtex4-1}
\usepackage{graphicx}
\usepackage{longtable} 
\usepackage{multirow}
\usepackage{hhline}
\usepackage{hyperref}
\usepackage{amsmath}
\usepackage{gensymb}
\usepackage{nicefrac}
\usepackage{amsmath}
\usepackage{hyperref}
\usepackage{color}
\usepackage{amssymb}
\usepackage{amsthm}
\usepackage{siunitx}
\usepackage{textcomp}
\begin{document}

\title{Investigation of spin-1 honeycomb antiferromagnet BaNi$_2$V$_2$O$_8$ with easy-plane anisotropy}
\author{E. S. Klyushina}
\affiliation{Helmholtz-Zentrum Berlin f{\"u}r Materialien und Energie, 14109 Berlin, Germany} 
\affiliation{Institut f{\"u}r Festk{\"o}rperphysik, Technische Universit{\"a}t Berlin, 10623 Berlin, Germany} 
\author{B. Lake}
\affiliation{Helmholtz-Zentrum Berlin f{\"u}r Materialien und Energie, 14109 Berlin, Germany} 
\affiliation{Institut f{\"u}r Festk{\"o}rperphysik, Technische Universit{\"a}t Berlin, 10623 Berlin, Germany} 
\author{A.T.M.N. Islam}
\affiliation{Helmholtz-Zentrum Berlin f{\"u}r Materialien und Energie, 14109 Berlin, Germany} 
\author{J. T. Park}
\affiliation{Heinz Maier-Leibnitz Zentrum, TU M{\"u}nchen, D-85747 Garching, Germany}
\author{A. Schneidewind}
\affiliation{Forschungszentrum J{\"u}lich GmbH, J{\"u}lich Centre for Neutron Science at MLZ, Lichtenbergstr. 1, D-85747 Garching, Germany} 
\author{T. Guidi}
\affiliation{ISIS Facility, Rutherford Appleton Laboratory, Chilton, Didcot, Oxon OX11 0QX, United Kingdom} 
\author{E. A. Goremychkin} 
\affiliation{ISIS Facility, Rutherford Appleton Laboratory, Chilton, Didcot, Oxon OX11 0QX, United Kingdom} 
\author{B. Klemke}
\affiliation{Helmholtz-Zentrum Berlin f{\"u}r Materialien und Energie, 14109 Berlin, Germany} 
\author{M. M{\aa}nsson}
\affiliation{Materials Physics, KTH Royal Institute of Technology, SE-16440 Stockholm Kista, Sweden} 
\affiliation{Laboratory for Neutron Scattering and Imaging, Paul Scherrer Institute, CH-5232 Villigen, Switzerland} 

\begin{abstract}
The magnetic properties of the two-dimensional, S=1 honeycomb antiferromagnet BaNi$_2$V$_2$O$_8$ have been comprehensively studied using DC susceptibility measurements and inelastic neutron scattering techniques. The magnetic excitation spectrum is found to be dispersionless within experimental resolution between the honeycomb layers, while it disperses strongly within the honeycomb plane where it consists of two gapped spin-wave modes. The magnetic excitations are compared to linear spin-wave theory allowing the Hamiltonian to be determined. The first- and second-neighbour magnetic exchange interactions are antiferromagnetic and lie within the ranges 10.90meV$\le$J$_n$$\le$13.35 meV and 0.85meV$\le$J$_{nn}$$\le$1.65 meV respectively. The interplane coupling J$_{out}$ is four orders of magnitude weaker than the intraplane interactions, confirming the highly two-dimensional magnetic behaviour of this compound. The sizes of the energy gaps are used to extract the magnetic anisotropies and reveal substantial easy-plane anisotropy and a very weak in-plane easy-axis anisotropy. Together these results reveal that BaNi$_2$V$_2$O$_8$ is a candidate compound for the investigation of vortex excitations and Berezinsky-Kosterliz-Thouless phenomenona.
\end{abstract}

%\pacs{}
%\keywords{}

% References should be done using the \cite, \ref, and \label command
\maketitle
\section{Introduction}
\par It is possible to realize low dimensional magnetic systems in real three-dimensional (3D) crystal structures. This is because in some cases the exchange interactions between magnetic ions exist only along particular directions within the crystal lattice due to their strong dependence on the overlap of electronic orbitals. Thus, the properties of low-dimensional magnetic systems can be investigated experimentally allowing the intriguing phenomena predicted for these Hamiltonians to be tested.
\par While physical realizations of numerous low-dimensional magnetic models such as one-dimensional (1D) spin-1/2 antiferromagnetic chains \cite{KCuF1Dchain}, Haldane chains \cite{Haldanechain} and spin-ladders \cite{spinladder} are already discovered and well explored, there are still models whose solid state prototypes are very rare or as yet undiscovered. A remarkable example is the two-dimensional (2D) Heisenberg magnet with easy-plane anisotropy which has attracted attention as a candidate to display Berezinsky\textendash Kosterliz \textendash Thouless (BKT) phenomenon \cite{BerezinskiiP1, BerezinskiiP2, KosterlitzThouless, Kosterlitz}.  
\par Although the existence of conventional long-range magnetic order at finite temperatures is prohibited by the Merim\textendash Wagner theorem \cite{Mermin} for 2D XY magnetic systems, a topological  phase transition from a disordered to a quasi-long-range magnetically ordered state has been theoretically proposed. Kosterliz and Thouless\cite{KosterlitzThouless, Kosterlitz} independently with Berezinsky \cite{BerezinskiiP1, BerezinskiiP2} suggested the presence of topological defects known as spin-vortices characterized by a topological winding number in 2D XY magnetic systems at finite temperatures. These vorticies couple into vortex/antivortex pairs below the finite transition temperature T$_{BKT}$ resulting in a quasi-long-range magnetically ordered state.
\par So far, the BKT transition was experimentally observed only in superfluids \cite{BKTsuperfluid} and superconducting thin films \cite{ThinFilms, ThinFilms2} which are good physical realizations of 2D XY models, or under applied magnetic field in Cu-based  dimerized coordination polymers \cite{BKTfield}. However, quantum Monte Carlo simulations \cite{Cuccoli,Cuccoli2} predict that the BKT transition can also be observed in 2D Heisenberg magnetic systems which have only weak easy-plane (XXZ-type) anisotropy spurring interest to search for and investigate magnetic compounds which are physical realization of the 2D XXZ Heisenberg model. 
\par Layered honeycomb antiferromagnets are promising candidates to realize 2D magnetic systems. Some indications of the free vortex phase have been found in the honeycomb antiferromagnet MnPS$_3$ where the temperature dependence of the magnetic correlation length revealed the predicted BKT exponential dependence rather than the usual power law dependence of conventional magnets \cite{Wildes2006,RonnowMnPS3}. However, the BKT transition is obscured in this compound by the onset of long-range magnetic order which occurs at a higher temperature. Two other well-explored honeycomb antiferromagnets are BaNi$_2$(AsO$_4$)$_2$ and BaNi$_2$(PO$_4$)$_2$ where the magnetic Ni$^{2+}$ ions have spin-1 (S=1) and also develop conventional long-range antiferromagnetically ordered ground states with easy-plane anisotropy \cite{Regnault3,Regnault2,Jongh}. The interplane magnetic exchange coupling in both compounds is extremely weak and was found to be three orders of magnitude weaker than the dominant intraplane magnetic exchange interaction. The heat capacity \cite{Regnault3}, however, revealed the behavior typical of a 3D Heisenberg magnetic system with a characteristic sharp $\lambda$-peak at a finite ordering temperature T$_N$, rather than the broad maximum and smooth features expected for the heat capacity of a 2D magnetic system\cite{Specificheat2D}.
\par The compound BaNi$_2$V$_2$O$_8$, belongs to the same class of Ni-based layered honeycomb antiferromagnets as BaNi$_2$(AsO$_4$)$_2$  and  BaNi$_2$(PO$_4$)$_2$, but has been much less explored. The crystal structure of BaNi$_2$V$_2$O$_8$ has space group R$\bar{3}$ with lattice constants a=b=5.0575 $\AA$ and c=22.33 $\AA$.  The S=1 magnetic Ni$^{2+}$ ions are surrounded by O$^{2-}$ ions forming edge-sharing NiO$_6$ octahedra which are arranged into honeycomb layers within the $ab$-plane. These layers are well separated along the c-axis by Ba$^{2+}$ ions and non-magnetic V$^{5+}$O$_4$ tetrahedra. The distance between the layers is three times larger than the distance between the nearest neighbor magnetic ions within the plane suggesting that the interactions between the honeycomb layers are weak. 
\par Powder neutron diffraction measurements reveal that the spins develop long-range magnetic order below the transition temperature T$_N$$\sim$50K \cite{Rogado}. In contrast, the reported DC magnetic susceptibility and heat capacity data display no clear signatures of the transition to the ordered state and suggest predominantly two-dimensional behavior. In particular, the heat capacity data \cite{Knafo} is in good agreement with the theoretical predictions of quantum Monte Carlo simulations performed for quasi-2D Heisenberg systems\cite{Specificheat2D} confirming the predominantly 2D character of BaNi$_2$V$_2$O$_8$. The presence of easy-plane anisotropy is revealed by the susceptibility data that shows a strong difference between the measurements with magnetic field applied in-plane and out-of-plane which persists to temperatures well above T$_N$ \cite{Rogado}. 
\begin{figure}
\includegraphics[width=1\linewidth]{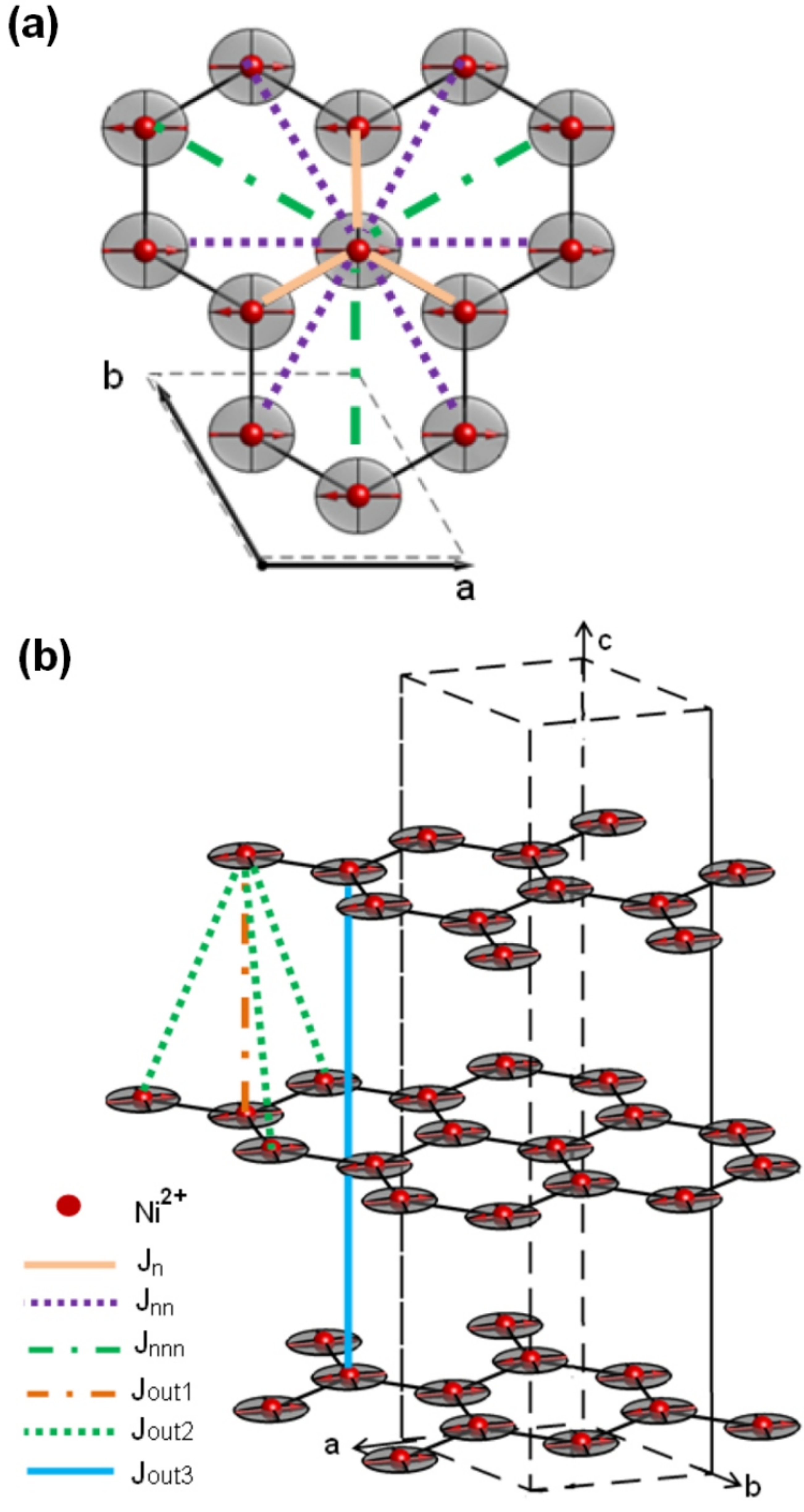} 
\caption{Crystal and magnetic structure (one twin) of BaNi$_2$V$_2$O$_8$ within (a) a honeycomb plane and (b) a single unit cell. Only the magnetic Ni$^{2+}$ ions are shown for clarity. The magnetic structure is based on the results of powder neutron diffraction measurements \cite{Rogado}. The red arrows represent spin directions and the gray ellipsoids illustrate the easy-plane single-ion anisotropy. The magnetic exchange interactions are represented by the colored lines and are discussed in Section III.C. (page \pageref{subsec:SecC}).} 
\label{fig:Fig1} 
\end{figure}
\par The magnetic structure of BaNi$_2$V$_2$O$_8$ was solved by Rogato {\em et.\ al.\ } \cite{Rogado}. In the ground state the spins lie within the honeycomb plane and are collinear so that each spin is antiparallel to its three nearest neighbours (red arrows in Fig.\ref{fig:Fig1}). This magnetic structure suggests that the dominant exchange interaction is antiferromagnetic between first nearest neighbor Ni$^{2+}$ ions (beige solid lines in Fig.\ref{fig:Fig1} (a)). Its strength was estimated to be J$_n$=8.27 meV from suseptibilty measurements \cite{Rogado} (all magnetic exchange constants are stated assuming single counting of each magnetic bond and where a positive sign defines an AFM interaction). Second neighbor interactions J$_{nn}$ (dotted magenta lines in Fig.\ref{fig:Fig1} (a)) and third-neighbor interactions J$_{nnn}$ (dashed-dotted green lines in Fig.\ref{fig:Fig1} (a)) may also be present. Figure \ref{fig:Fig1} (b) shows the possible magnetic exchange paths responsible for the  interplane coupling. J$_{out1}$ (dashed-dotted orange lines) and J$_{out2}$ (dotted green lines) correspond to the shortest and next-shortest distances between parallel and antiparallel spins in neighboring honeycomb planes respectively, and the path J$_{out3}$ is realized via the shortest distance between antiparallel spins in the first and third honeycomb layers. 
\par Altogether the available results for BaNi$_2$V$_2$O$_8$  imply that this compound is a quasi-2D Heisenberg antiferromagnet with easy-plane anisotropy and suggest it is suitable for investigations of BKT phenomena. Indeed, recent $^{51}$V nuclear magnetic resonance\cite{Heinrich} and electronic spin resonance\cite{Waibel2015} measurements reveal that the correlation length of BaNi$_2$V$_2$O$_8$ decays exponentially as a function of temperature in agreement with the Berezisky-Kosterlitz-Thouless theory\cite{Kosterlitz}.
\par Despite these promising results, the Hamiltonian of BaNi$_2$V$_2$O$_8$ is currently unknown.  In this paper the magnetic properties of this compound are explored using DC susceptibility measurements along with powder and single crystal inelastic neutron scattering (INS) measurements. The DC susceptibility measurements are performed up to high temperatures ($2 K \le$T$\le 640$ K) which allow the total value of the magnetic exchange interactions, J$_{total}$, to be determined. The single crystal INS measurements at base temperature reveal spin-wave excitations which disperse within the honeycomb plane and are completely dispersionless in the out-of-plane direction confirming the strongly 2D magnetic behavior of BaNi$_2$V$_2$O$_8$. The magnetic excitation spectrum consists of two modes which are both gapped suggesting a dominant easy-plane anisotropy and a weak easy-axis anisotropy which gives the prefered directions within this plane. The magnetic excitations are analyzed using linear spin-wave theory which allows the exchange interactions and anisotropies to be extracted. The Hamiltonian of BaNi$_2$V$_2$O$_8$ is solved and found to be closer to an ideal 2D antiferromagnet with easy-plane anisotropy than BaNi$_2$(AsO$_4$)$_2$ and BaNi$_2$(PO$_4$)$_2$. Thus, this compound is a promising candidate to display BKT phenomena above T$_N$ where the system is in a quasi-2D XY regime. According to the BKT theory\cite{KosterlitzThouless, Kosterlitz} this behaviour manifests itself as an exponential decay of the order parameter above the transition temperature T$_{BKT}$.
\section{Experimental details}
\par Powder and single crystalline samples of BaNi$_2$V$_2$O$_8$ were grown in the Core Lab for Quantum Materials at the Helmholtz Zentrum Berlin f{\"u}r Materialien und Energie (HZB) in Germany. The single crystal growth was performed in a four mirror optical image furnace using the Traveling-solvent-floating-zone method. Feed rods for the growth were prepared by solid state reaction of high purity powders of BaCO$_3$, NiO and V$_2$O$_5$ in air. A solvent with excess BaV$_2$O$_6$ was attached to the tip of the feed rod as a flux compound to facilitate the growth at a lower temperature. The growth was performed at a rate of 0.2 mm/h in air atmosphere. 
\par Single crystal static magnetic susceptibility measurements were performed on BaNi$_2$V$_2$O$_8$ at the Core Lab for Quantum Materials, HZB, using a Physical Property Measurement System (PPMS) and a Magnetic Property Measurement System with a Vibrating Sample Magnetometer (MPMS-VSM) by Quantum Design. Three single crystal samples with masses of 8.48 mg, 19.38 mg and 4.85 mg were cut from different single crystal growths and data were collected under a magnetic field of 1T applied both parallel and perpendicular to the c-axis. The measurements were performed at both low (2~-~400~K) and high (300~-~640~K using a heater stick) temperatures. The upper temperature was limited to be below the temperature of 650 K at which the sample undergoes structural changes. The data collected in the low- and high-temperature regimes were combined by averaging over the overlapping temperature range (300-400 K).   
\par Both powder and single crystal inelastic neutron scattering (INS) measurements were performed to explore the magnetic excitation spectrum of BaNi$_2$V$_2$O$_8$. Powder INS data were collected using a new high count rate thermal time-of-flight spectrometer MERLIN \cite{Merlin} at ISIS, Rutherford Appleton Laboratory, U.K. The  polycrystalline sample with total mass of 19.03g was cooled to the base temperature of 5K using a closed cycle refrigerator. The Fermi chopper was operated in four different modes phased to give incident neutron energies of E$_i$=80 meV, E$_i$=50 meV, E$_i$=35 meV and E$_i$=20 meV respectively. For these energies the chopper frequencies were 400Hz, 300Hz, 250Hz and 200Hz and the corresponding resolutions at the elastic energy were E$_i$=5.1$\pm$0.1 meV, E$_i$=3.2$\pm$0.06 meV, E$_i$=2.26$\pm$0.06~meV and E$_i$=1.24$\pm$0.02 meV, respectively. These resolution values were extracted by fitting the incoherent elastic signal. A vanadium standard was used to normalize the detectors.
\par Single crystal INS measurements were performed on the thermal neutron triple-axis spectrometer PUMA\cite{Puma} at the Heinz Maier-Leibnitz Zentrum (MLZ), Garching, Germany. The single crystal sample of BaNi$_2$V$_2$O$_8$ with mass of 340mg was preoriented with the (h,k,0) plane as the horizontal instrumental scattering plane and was cooled to the base temperature of 3.5 K by a closed-cycle cryostat. The instrument was equipped with a double-focused pyrolytic graphite (PG(002)) monochromator and analyzer operated at the fixed final wave-vector of k$_f$ =2.662\AA$^{-1}$ giving an elastic energy resolution of 0.8 meV. A sapphire filter was located before the monochromator to reduce the fast neutron background while two PG filters were placed in series between the sample and analyser to eliminate higher-order neutrons from the scattered beam.
\par For a detailed investigation of the low energy excitation spectra, the same sample was measured on the cold neutron triple-axis spectrometer PANDA at the Heinz Maier-Leibnitz Zentrum (MLZ), Garching, Germany. PANDA was equipped with a double-focusing PG(002) monochromator and analyzer. The measurements were performed in both the (h,k,0) and (h,0,$l$) scattering planes at the base temperature of 4.3K, which was achieved by using a closed-cycle cryostat.  All data were collected at the fixed final wave vector k$_f$= 1.57 \AA$^{-1}$ and the elastic energy resolution was 0.1379 meV. A Beryllium filter was used to remove the high-order wavelengths diffracted by the monochromator.
\par A larger single crystal of BaNi$_2$V$_2$O$_8$ with mass of 500mg was measured using the Cold Neutron Triple-Axis spectrometer TASP \cite{Tasp}, at the Paul Scherrer Institute (PSI), Switzerland. The sample was preoriented with (h,k,0) plane as the instrumental scattering plane and was cooled down to 1.47K using an orange cryostat. 
\par The measurements were carried out in the (h,k,0) and (h,k,h/2) scattering planes where (h,k,h/2) was achieved by tilting (h,k,0) by 5$\degree$ about the horizontal axis perpendicular to the (0,1,0) direction. This instrument was equipped with a vertically focused PG(002) monochromator and a horizontally focused PG(002) analyzer. The final wave vector was fixed at k$_f$ =1.23\AA$^{-1}$ giving an energy resolutions of 0.065 meV. 
\par To further explore the magnetic excitation spectra of BaNi$_2$V$_2$O$_8$ along all reciprocal space directions including off-symmetry directions, the single crystalline sample of BaNi$_2$V$_2$O$_8$ with mass of 500mg was measured using the cold-neutron multichopper spectrometer LET\cite{LET} at ISIS, Rutherford Appleton Laboratory, U.K. This spectrometer uses the time-of-flight technique which allows a large region of reciprocal space (several Brillouin zones) and a wide range of energy transfers to be measured simultaneously. The orientation of the sample was aligned with the (h,k,0) plane as the horizontal instrument scattering plane and cooled down to T=5K using a closed cycle refrigerator. The INS data were collected at four different incident energies of E$_{i}$=2.97 meV,  E$_{i}$=5 meV, E$_{i}$=10.2 meV and E$_{i}$=30 meV simultaneously which are characterized by calculated elastic energy resolutions of $\Delta$E$_{i}$=0.05 meV,  $\Delta$E$_{i}$=0.1 meV, $\Delta$E$_{i}$=0.31 meV and $\Delta$E$_{i}$=1.2 meV respectively.    
\par The powder and single crystal magnetic excitation spectra of BaNi$_2$V$_2$O$_8$ were analysed using the SpinW Matlab Library \cite{SpinW} which can perform numerical calculations of the spin-wave energies and intensities for complex magnetic lattices with commensurate or incommensurate magnetic order.
\begin{figure*}
\includegraphics[width=1\linewidth]{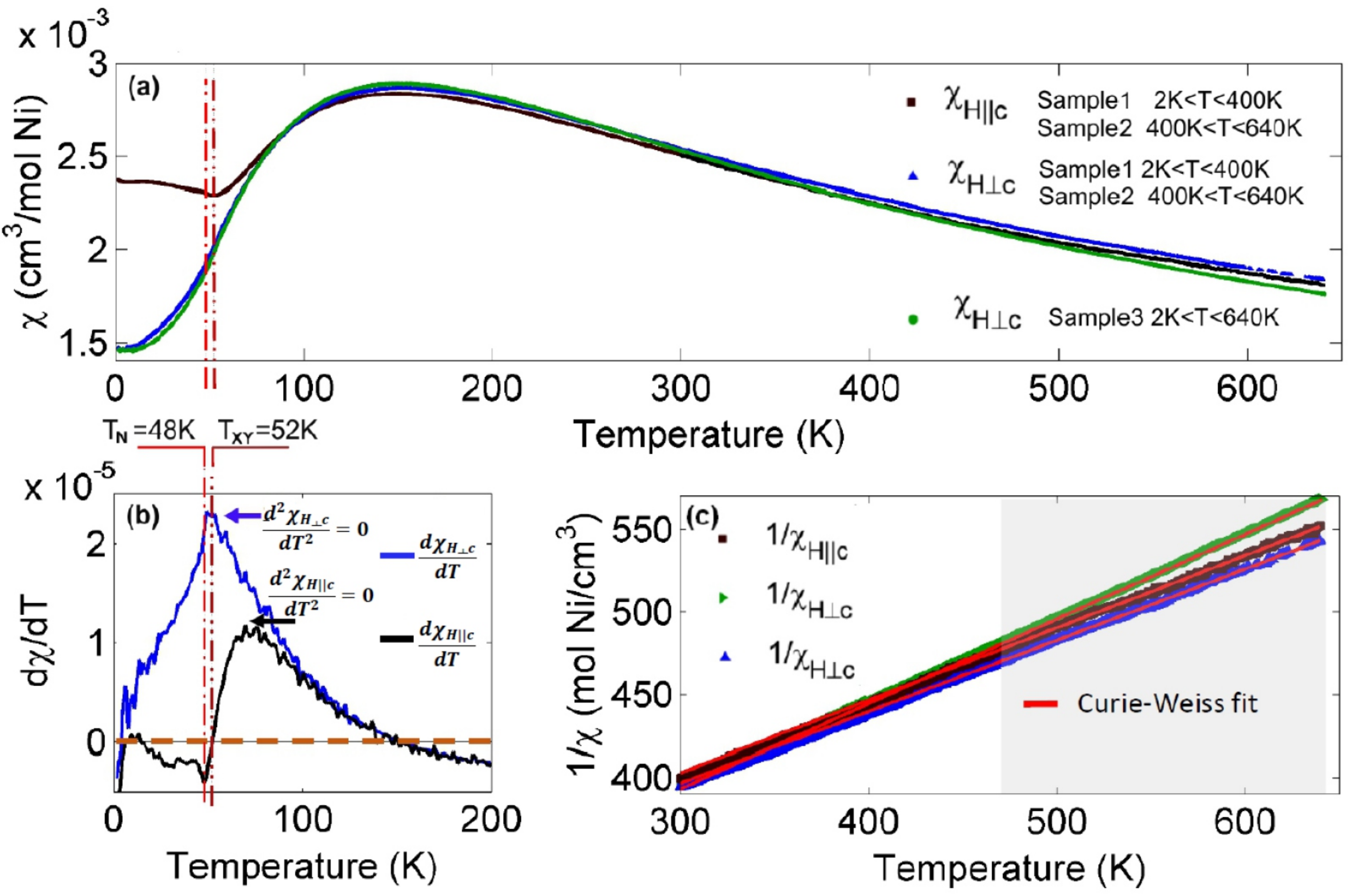} 
\caption{(a) Single crystal DC magnetic susceptibility data of BaNi$_2$V$_2$O$_8$ collected under a magnetic field of 1T applied parallel and perpendicular with respect to the c-axis. A Curie-impurity tail was subtracted from the data collected on sample 3. (b) The first derivative of $\chi_{H||c}$ and $\chi_{H\perp c}$ with respect to temperature.  (c) The inverse susceptibilities ($\chi_{H||c}^{-1}$ and $\chi_{H\perp c}^{-1}$) plotted as a function of temperature and fitted using the Curie-Weiss law as described in the text. The fitted region lies within the shaded area.} 
\label{fig:Fig2}
\end{figure*}  
\section{Results and Analysis}
\subsection{Static Susceptibility}      
\par Figure \ref{fig:Fig2}(a)  shows the single crystal DC magnetic susceptibility data collected on the single crystalline samples of BaNi$_2$V$_2$O$_8$ over the temperature range 2K to 640K under a DC magnetic field of 1T applied parallel ($\chi_{H||c}$) and perpendicular ($\chi_{H\perp c}$) to the c-axis. The data were collected on different single crystalline samples to improve the statistics and to check for consistency. 
\par Both $\chi_{H||c}$ and $\chi_{H\perp c}$ display a broad maximum in the vicinity of T$\approx$150K which indicates the presence of short-range magnetic correlations.  Above this maximum the susceptibilities show the same behavior and smoothly decrease with increasing temperature suggesting that the magnetism is isotropic. A small difference between the magnitude of $\chi_{H||c}$ and $\chi_{H\perp c}$ observed at high temperatures can be attributed to the anisotropy of the  Land{\'e} {\it{g}}-factor which has the values {\it{g}$_{H\perp c}$}=2.243 and {\it{g}$_{H||c}$}=2.225 \cite{Heinrich}. 
\par Below the maximum the susceptibility starts to display anisotropic behavior which first appears at T$_{ani}$=80K below which $\chi_{H||c}$ and $\chi_{H\perp c}$ split from each other. The anisotropy becomes more pronounced as temperature decreases. Indeed, below 80K, $\chi_{\perp c}$ continuously decreases, in contrast to $\chi_{H||c}$  which shows a weak minimum at T$_{XY}$=52 K and then increases slightly and becomes constant with further temperature decrease. 
\subsubsection{Low Temperature Susceptibility} 
\par Figure \ref{fig:Fig2}(b) shows the first derivative of $\chi_{H||c}$ and $\chi_{H\perp c}$ with respect to temperature plotted over the temperature range from 2K to 200K. This plot reveals that at T=52K $\frac{d\chi_{H||c}}{dT}$=0, satisfying the condition for an extremum and confirming that $\chi_{H||c}$ has a minimum at this temperature. The first derivative of $\chi_{H\perp c}$ with respect to temperature has a sharp maximum at T=52K which means that second derivative satisfies the condition for an inflection point ($\frac{d^2\chi_{H\perp c}}{dT^2}$=0) at this temperature. 
\par These results can be compared with the predictions of Quantum Monte Carlo (QMC) simulations performed by Cuccoli et. al.\cite{Cuccoli,Cuccoli2} for the 2D, spin-1/2 square lattice with Ising (XXZ-type) interactions. According to these simulations the in-plane susceptibility decreases continuously with decreasing temperature \cite{Cuccoli,Cuccoli2}. In contrast, the out-of-plane susceptibility has a minimum at a finite temperature T$_{co}$ which was assigned by Cuccoli et. al.\cite{Cuccoli,Cuccoli2} to the transition to a regime where the spins lie predominantly in the XY plane. Thus, the minimum observed in the $\chi_{H||c}$ data of BaNi$_2$V$_2$O$_8$ at T$_{XY}$=52K could be a signature of a crossover to a regime dominated by easy-plane anisotropy.
\par Indeed, the behavior observed in $\chi_{H||c}$ and $\chi_{H\perp c}$ in BaNi$_2$V$_2$O$_8$ below T$_{XY}$=52K is consistent with the behavior expected for $\chi_{H||c}$ and $\chi_{H\perp c}$ of a conventional AFM below its Neel temperature T$_N$ (when the spins are ordered within the plane perpendicular to the c-axis). However, the temperature T$_{XY}$=52K is higher than T$_N$=50K proposed in the literature for BaNi$_2$V$_2$O$_8$ \cite{Rogado}. 
\par Neither $\chi_{H||c}$ nor $\chi_{H\perp c}$ reveal clear signatures for the phase transition to the magnetically ordered state proposed at $\approx$50K by powder neutron diffraction measurements\cite{Rogado}. Indeed, only the first derivative of the $\chi_{H||c}$ has a sharp minimum at 48K$\pm$0.5K satisfying the condition $\frac{d^2\chi_{H||c}}{dT^2}$=0 for an inflection point which is in agreement with the tiny feature observed at the same temperature in heat capacity data\cite{Knafo} and suggests that the phase transition to the long-range magnetically ordered state occurs at T$_N$=48K$\pm$0.5K. 
\subsubsection{Curie-Weiss Law}   
\par The magnetic susceptibility data were analyzed using the Curie-Weiss law $\chi(T)$=$\nicefrac{C}{(T-\Theta_{CW})}$ where $C$ is the Curie constant and $\Theta_{CW}$ is the Curie-Weiss temperature. Figure \ref{fig:Fig2}(c) shows the inverse magnetic susceptibilities plotted as a function of temperature over the range 300K-640K, the data is almost linear suggesting that BaNi$_2$V$_2$O$_8$ is in its paramagnetic regime. The Curie-Weiss law was fitted to each dataset over the temperature range 470K-640K (solid red line in Fig.\ref{fig:Fig2}(c)) and the extracted values of the Curie-Weiss temperature and Curie constant were averaged over the different samples and directions of the applied magnetic field. The Curie-Weiss temperature and  Curie constant of BaNi$_2$V$_2$O$_8$ were found to be $\Theta_{CW}$= -542$\pm$76K and $C$=1.06$\pm$0.12cm$^3$K/mol, respectively. The negative sign of the Curie-Weiss temperature confirms that the dominant magnetic exchange interaction in BaNi$_2$V$_2$O$_8$ is antiferromagnetic (AFM). 
\\* The Curie-Weiss temperature can be used to estimate the sum of the magnetic exchange interactions up to the third-neighbor interaction in BaNi$_2$V$_2$O$_8$ using the relation \cite{Choi}
\begin{equation}\label{eq:Jtotal}
{{\Theta_{CW}}= -\frac{S(S+1)(g-1)^2\sum_{i=n,nn,nnn}z_{i}J_{i}}{3k_{B}}}
\end{equation}
Here, S=1 is the spin value, k$_{B}$ is the Boltzmann constant, $g$ is the Land{\'e}-factor and z$_{i}$ is the number of neighbors coupled to each magnetic ion by the magnetic exchange interactions J$_{i}$. In the above relation each interaction J$_{i}$ is counted only once per bond. In BaNi$_2$V$_2$O$_8$ each Ni$^{2+}$ magnetic ion has three first nearest neighbors, six second-nearest neighbors and three third-nearest neighbors (Fig.\ref{fig:Fig1} (a)), therefore, $\sum_{i=n,nn,nnn} z_{i}J_{i}~=~3J_{n}+6J_{nn}+3J_{nnn}$=3J$_{total}$ where J$_{total}$ is defined as J$_{total}\equiv J_{n}+2J_{nn}+J_{nnn}$. Substituting the extracted value of $\Theta_{CW}$= -542$\pm$76K into Eq.\ \eqref{eq:Jtotal} yields the value J$_{total}$=15$\pm$2 meV. 
\begin{figure*} 
\includegraphics[width=0.85\linewidth]{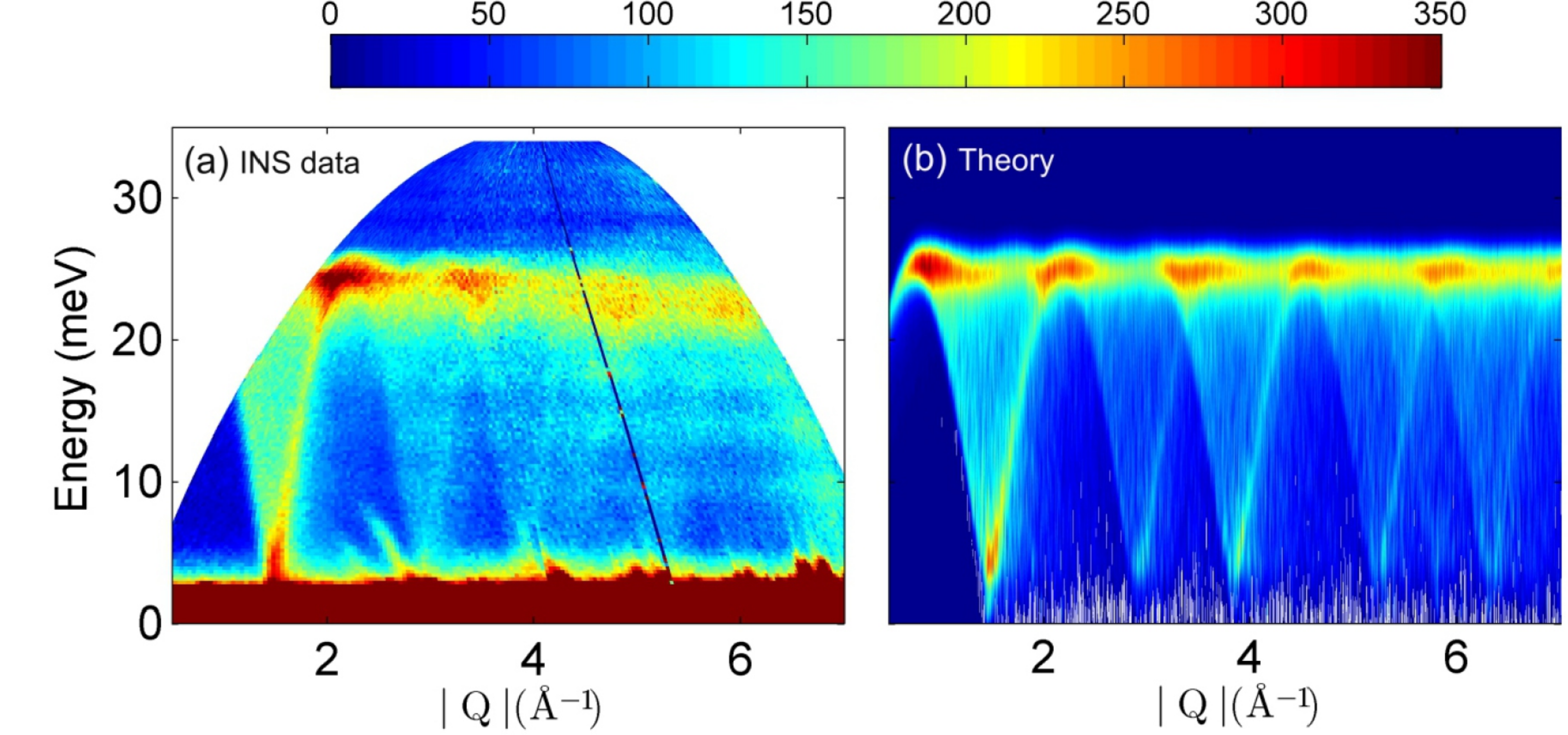} 
\caption{ (a) INS data measured on a powder sample at T=5K using the Merlin spectrometer with an incident energy of 35 meV. (b) Spin-wave simulations of the powder excitation spectrum of BaNi$_2$V$_2$O$_8$ performed for the Hamiltonian eq.\eqref{eq:Hamiltonian} with the parameters J$_n$=12.3 meV, J$_{nn}$=1.25 meV, J$_{nnn}$=0.2 meV, J$_{out}$=-0.00045 meV, D$_{EP}$=0.0695 meV, D$_{EA}$=-0.0009 meV.}
\label{fig:Fig3} 
\end{figure*} 
\subsubsection{Quenching of the orbital angular momentum}
\par The Curie constant can be defined as $C$=$\nicefrac{N_A\cdot\mu^2_B\mu^2_{eff}}{3k_B}$ where $N_A$ is the Avogadro number, $\mu_B$ is the Bohr magneton, $k_B$ is the Boltzmann constant and $\mu_{eff}$ is the effective moment of the magnetic ions. Using the value of the Curie constant $C$=1.06$\pm$0.12cm$^3$K/mol, extracted from the magnetic susceptibility data, the effective magnetic moment of the Ni$^{2+}$ magnetic ions in BaNi$_2$V$_2$O$_8$ is found to be $\mu_{eff}$=2.91$\pm$0.16 $\mu_B$/Ni$^{2+}$. This value is much lower than the theoretical value of $\mu_{eff}$ =5.59 $\mu_B$/Ni$^{2+}$ calculated for the 3d$^8$ electronic configuration of Ni$^{2+}$ assuming unquenched orbital angular momentum, but is in good agreement with the value $\mu_{eff}$ =2.83 $\mu_B$/Ni$^{2+}$ expected for the electronic configuration where the orbital angular momentum is quenched (L=0, J=S=1, g$_J$=2). Such quenching is expected and commonly observed for the 3d ions in crystalline environments due to strong crystal field effects \cite{Blundell}. 
\subsection{Magnetic excitation spectrum of BaNi$_2$V$_2$O$_8$} 
\subsubsection{Powder magnetic excitation spectum}
\par The magnetic excitations of  BaNi$_2$V$_2$O$_8$ were explored using inelastic neutron scattering. First, to get an overview of the magnetic excitation spectrum and to estimate its energy range, INS data were collected on a powder sample at 5K using the MERLIN time-of-flight spectrometer. The results are plotted on Fig.\ref{fig:Fig3}(a) as a function of energy- and wavevector transfer, and reveal V-shaped spin-wave type excitations which extend over the energy range 0-26 meV. The excitations have their first minimum at $\mid{Q}\mid$=1.45\AA$^{-1}$ which corresponds to the proposed (1,0,$\frac{1}{2}$) magnetic Bragg peak position. A Van Hove singularity is observed at 26 meV indicating the top of energy dispersion. The intensity of the excitations is strong at low wavevectors and decreases with increasing wavevector confirming their magnetic origins.
\subsubsection{Single crystal magnetic excitation spectrum within the honeycomb plane} \label{sssec:SecB2} 
\begin{figure*} 
\includegraphics[height=0.9\textheight]{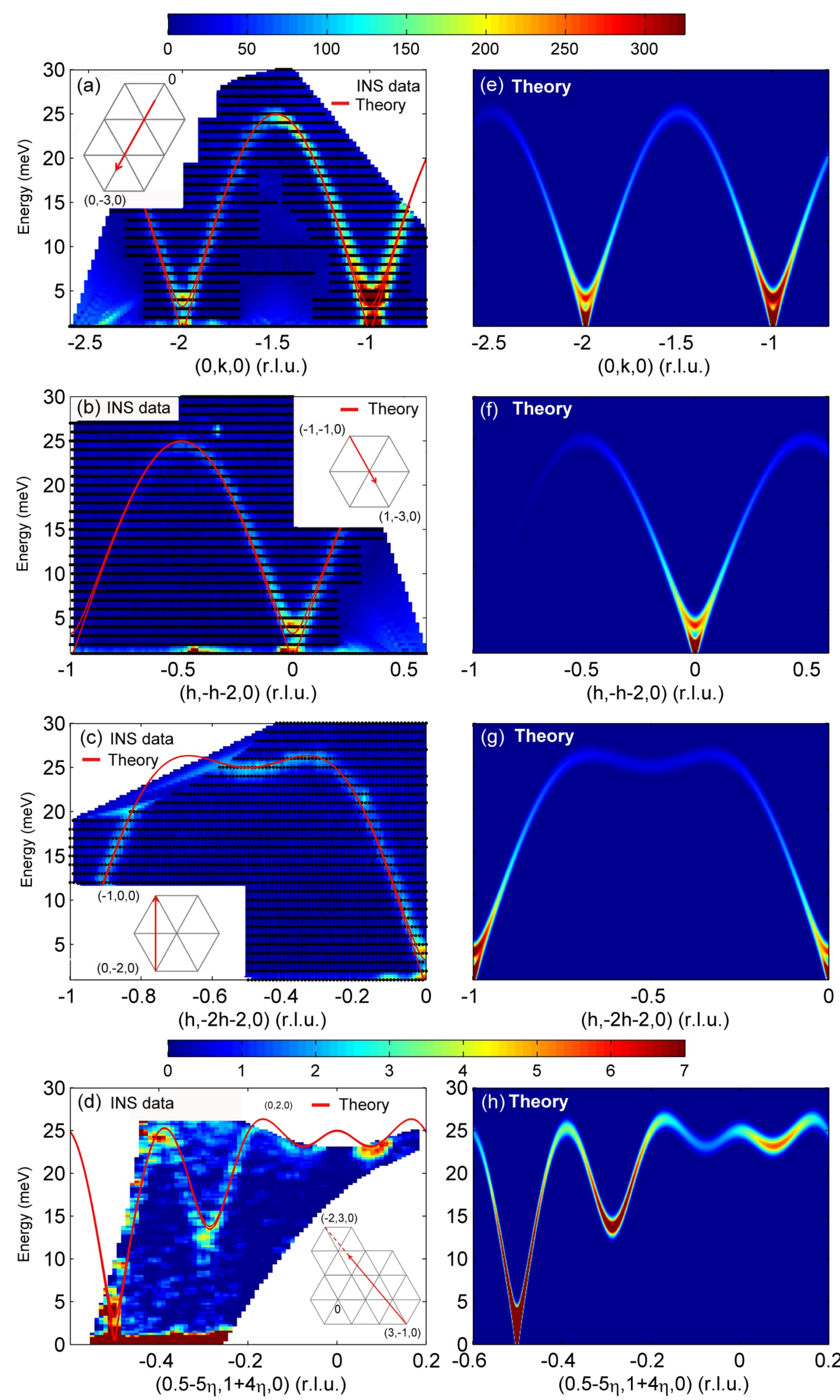} 
\caption{Single crystal INS data and spin-wave simulations of BaNi$_2$V$_2$O$_8$ plotted as a function of energy and wavevector transfer within the honeycomb plane. (a)-(c) INS data measured on the PUMA spectrometer at T=3.5K along the directions (a) (0,k,0)  (b) (h,-h-2,0) and (c) (h,-2h-2,0) as indicated by the red arrows in the insets. The black dotted lines indicate where the data were collected, features which appear outside these areas are artifacts and should be ignored. (d) Cut along the (0.5-5$\eta$,1+4$\eta$,0) direction through the INS data collected at T=5K on the cold-neutron multi-chopper LET spectrometer with incident energy of 30.2 meV. (e)-(h) Spin-wave simulations of the single crystal magnetic cross-section along the directions (e) (0,k,0), (f) (h,-h-2,0), (g) (h,-2h-2,0) and (h) (0.5-5$\eta$,1+4$\eta$,0). The calculated spin-wave energies are represented by the solid red lines through the data on (a)-(d). All simulations were performed for the Hamiltonian eq.\eqref{eq:Hamiltonian} with the parameters J$_n$=12.3~meV, J$_{nn}$=1.25 meV, J$_{nnn}$=0.2 meV, J$_{out}$=-0.00045 meV, D$_{EP}$=0.0695 meV, D$_{EA}$=-0.0009 meV.} 
\label{fig:Fig4} 
\end{figure*}
\begin{figure*}
\includegraphics[width=1.0\textwidth]{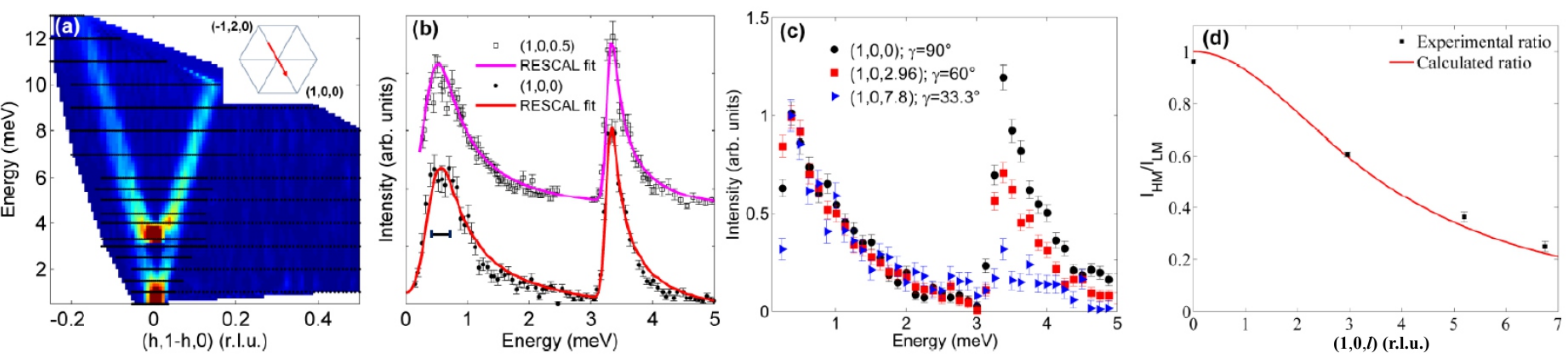} 
\caption[width=2\linewidth]{Low energy single crystal INS data of BaNi$_2$V$_2$O$_8$. (a) INS measured on the PANDA spectrometer at T=4.3K along the (h,1-h,0) direction, the black dotted lines indicate where the data were collected. (b) Energy scans at constant wave-vectors of (1,0,0) (filled black circles) and (1,0,0.5) (open black squares) collected on the TASP spectrometer at T=1.47K. The solid red and magenta lines represent the fit performed using the RESCAL software. The horizontal black lines represents the resolution broadening expected for a sharp mode. (c) Energy scans measured at T=4.3K using the PANDA spectrometer at (1,0,$l$) where $l$ is fixed to $l$=0 (filled black circles), $l$=2.96 (filled red squares) and $l$=7.8 (filled blue triangles). $\gamma$ is the angle between the wavector transfer $\vec{Q}$ and the c-axis. The background has been substracted and the intensities have been scaled so that the amplitude of the lower energy mode is fixed to 1.  (d) The ratio of the intensity of the higher energy mode to the lower energy mode extracted experimentally from the PANDA data shown in (c) (filled black squares) and calculated using the eq.\eqref{eq:ratio} (solid red line).} 
\label{fig:Fig5}
\end{figure*}  
\par Single crystal INS measurements of BaNi$_2$V$_2$O$_8$ were performed using the PUMA thermal triple-axis spectrometer to explore the magnetic excitation spectrum within the honeycomb plane. The INS measurements were carried out at T=3.5K as a function of energy and wavevector along different high symmetry directions within the (h,k,0) plane. Figure \ref{fig:Fig4}(a) shows the spectrum along the (0,k,0) direction which was constructed from a series of constant energy scans.  The results reveal that the magnetic excitations of BaNi$_2$V$_2$O$_8$ consist of two spin-wave modes which both disperse from the magnetic Bragg peak positions and are consistent with the honeycomb lattice. The two modes are degenerate except at low energies. The dispersion along the (h,-2-h,0) direction (Fig. \ref{fig:Fig4}(b)) is similar to that along the (0,k,0) direction (Fig. \ref{fig:Fig4}(a)) as expected since these directions are equivalent due to the lattice symmetry. Figure \ref{fig:Fig4}(c) shows that the dispersion along the (h,-2h-2,0) direction has a modulation in the high energy region which can be attributed to the presence of a weak second neighbour magnetic exchange interaction within the honeycomb plane as show later. 
\par To further explore the magnetic excitation spectra of BaNi$_2$V$_2$O$_8$ along all directions including off-symmetry directions, single crystal INS measurements were also performed on the high resolution cold-neutron multi-chopper LET spectrometer at the base temperature of T=5K. These results confirm the data which was collected on the PUMA spectrometer along specific high symmetry directions and, furthermore, provide additional data along low-symmetry directions such as that shown in Fig. \ref{fig:Fig4}(d).
\par To explore the low-energy  magnetic excitation spectrum of BaNi$_2$V$_2$O$_8$ in detail, single crystal INS measurements were carried out within the (h,k,0) scattering plane using the cold neutron PANDA spectrometer. High resolution constant energy scans were measured at T=4.3K along the (h,1-h,0) direction over the energy range of 0 - 13 meV with energy steps of 0.5 meV. The results are plotted in Fig. \ref{fig:Fig5}(a) and confirm the presence of two spin-wave modes dispersing from (0,1,0). A high resolution energy scan was performed at the dispersion minimum (1,0,0), at T=1.5K using the cold-neutron TASP spectrometer. The result is plotted in Fig.\ref{fig:Fig5}(b) and shows two peaks corresponding to the two spin-wave modes. The data clearly reveal that both modes are gapped. 
\par In order to obtain accurate values of these two energy gaps, the constant wavevector scan at the dispersion minimum (1,0,0) were fitted using the RESCAL software which convolves the instrumental resolution function calculated from the instrument settings with the dispersion relations. The dispersion relations \cite{Chernyshev2012} calculated for the Heisenberg honeycomb antiferromagnet within linear spin-wave theory were found to be appropriate to describe the magnetic excitations of BaNi$_2$V$_2$O$_8$. Energy gaps were introduced to the spin-wave dispersion relations\cite{Chernyshev2012} in order to reproduce the gaps observed experimentally. The fit to the energy scan at (1,0,0)  is given by the solid red line in Fig. \ref{fig:Fig5}(b) and provides good agreement with the experimental INS data. The low-energy mode is found to be substantially broader than the instrumental broadening (represented by the horizontal black line) due to the presence of the several split modes as will be shown later (see section III.C on page \pageref{subsec:SecC}). The higher energy mode is resolution limited. The fitted values of the energy gaps were found to be  E$_1$=0.41$\pm$0.03 meV and E$_2$=3.25$\pm$0.03~meV for the lower and higher energy modes respectively. 
\begin{figure*}
\includegraphics[width=1\textwidth]{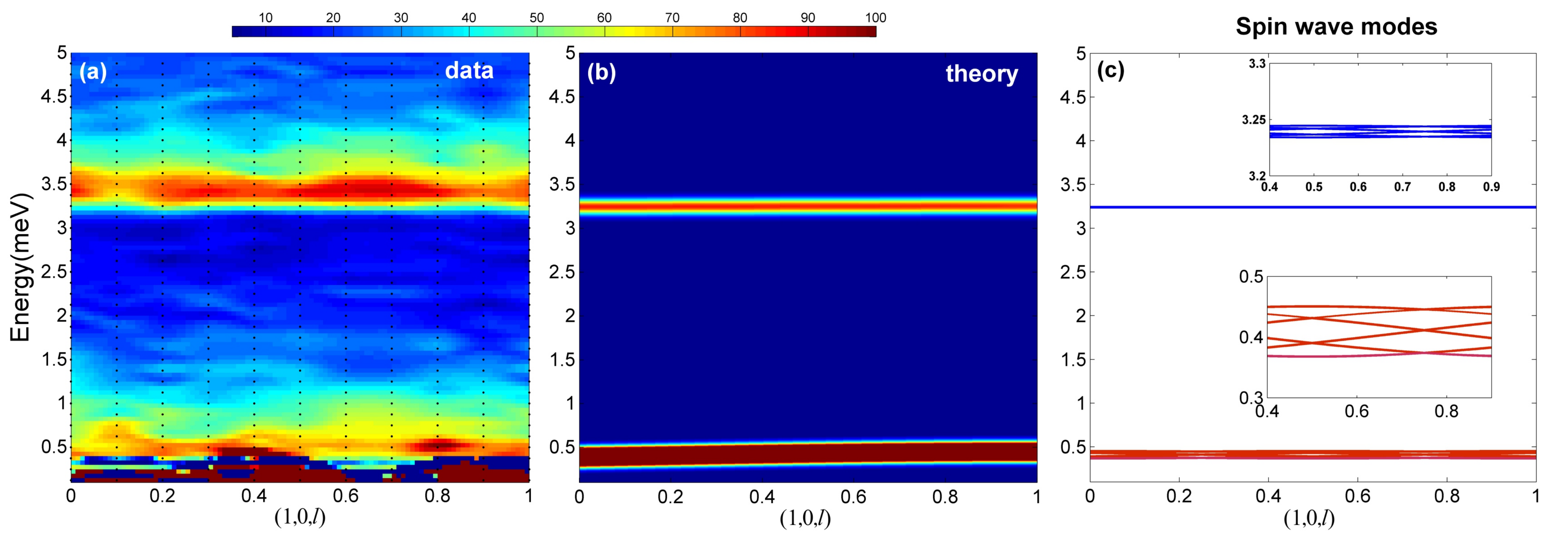} 
\caption[width=2\linewidth]{(a) Single crystal INS data of BaNi$_2$V$_2$O$_8$ measured on the PANDA spectrometer at T=4.3K. The constant wave-vector scans were performed at (1,0,$l$) over the energy range of 0-5 meV with $l$ fixed at values from 0 to 1 in step of 0.1 to cover one Brillouin zone. (b) Spin-wave simulations of the single crystal magnetic cross section along the (1,0,$l$) direction convolved with a Gaussian function. The simulations were performed for the Hamiltonian eq.\eqref{eq:Hamiltonian} with the parameters J$_n$=12.3~meV, J$_{nn}$=1.25~meV, J$_{nnn}$=0.2~meV, J$_{out}$=-0.00045~meV, D$_{EP}$=0.0695~meV, D$_{EA}$=-0.0009~meV. (c) The calculated dispersion relation of the twelve spin-wave modes along the (1,0,$l$) direction. (c.1)-(c.2) An enlarged representation of the dispersions of the six higher and six lower energy spin-wave modes, respectively.} 
\label{fig:Fig6}
\end{figure*} 
\subsubsection{Anisotropy of the spin fluctuations}\label{sssec:SecB3}
\par The presence of gapped magnetic excitations in the spin-wave spectra is often an indication of anisotropy because it suggests that there is a finite energy cost for the spins to fluctuate along specific crystallographic directions. The presence of two gapped modes at the in-plane antiferromagnetic-zone center suggests that BaNi$_2$V$_2$O$_8$ has both easy-plane and easy-axis anisotropies. One of the gapped spin-wave modes is probably due to spin fluctuations within the honeycomb plane perpendicular to the ordering direction which must overcome the easy-axis anisotropy, while the other is due to the fluctuations along the out-of-plane (c-axis) direction which must overcome the easy-plane anisotropy. Since the two modes have different gap sizes the two anisotropies have different strengths.
\par The intensity variation of the modes at their gap energies as a function of wave-vector transfer can be used to verify the type of spin-fluctuations to which they correspond. Indeed, the magnetic INS cross-section depends on the direction of the wavevector transfer $\vec{\bf{Q}}$ with respect to the direction of the spin fluctuations $\Delta\vec{\bf{S}}$ so that only the component of the spin fluctuations which are perpendicular to the wave-vector transfer will contribute to the neutron scattering cross-section. 
\par To explore the origins of the observed energy gaps, constant-wavevector scans were measured at (1,0,$l$) over the energy range 0 to 7 meV at different fixed values $l$=0, 2.96, 5.2, 6.75, 7.8. The data were collected at 4.3K using the PANDA spectrometer. Figure \ref{fig:Fig5}(c) shows the constant wavevector scans at (1,0,0), (1,0,2.96) and (1,0,7.8). To simplify the comparison each energy scan has been rescaled so that the intensity of the lower-energy mode is 1. The results reveal that the amplitude of the high-energy mode decreases as the wavevector transfer becomes increasingly parallel to the c-axis. Indeed, both modes have approximately the same intensities at (1,0,0), where the wavevector transfer lies entirely within the honeycomb plane. In contrast, for the energy scan at (1,0,7.8) where the wavevector transfer is mostly along the $l$-direction, the high energy mode almost disappears. These results suggest that the higher-energy mode is caused by out-of-plane spin fluctuations parallel to the c-axis while the lower energy mode is due to in-plane spin-fluctuations. 
\par To quantitatively analyse this observed wavevector dependence, the ratio of the intensity of the higher energy mode (I$_{HM}$) to the lower energy mode (I$_{LM}$) for the constant-wavevector scans at (1,0,0), (1,0,2.96), (1,0,5.2) and (1,0,6.75) was extracted and plotted as a function of $l$ in Fig.\ref{fig:Fig5}(d). The constant wave vector scan at (1,0,7.8) was not analysed because the higher energy mode is extremely weak. The result reveals that this ratio continuously decreases as $l$ increases from $l$=0 to $l$=6.75.   
\par The experimental wavevector dependence of the ratio $\frac{I_{HM}}{I_{LM}}$ was compared with the theoretical ratio $\frac{I_{c}}{I_{(a,b)}}$ of the intensities of the modes due out-of-plane ($I_{c}$) and in-plane ($I_{(a,b)}$) spin fluctuations for the magnetic structure of BaNi$_2$V$_2$O$_8$ taking account of the in-plane twinning (see Appendix):
\begin{equation}\label{eq:ratio}
{\frac{I_{c}}{I_{(a,b)}} \propto \frac{(1-cos^2(\gamma))}{(1-(\frac{1}{3}\cdot sin^2(\gamma)+ \frac{2}{3}\cdot sin^2(\gamma)\cdot cos^2(60)))}}
\end{equation} 
\par $\gamma$ is the angle between the wave-vector transfer $\bf{Q}$ and the c$^*$-axis and is given by the relation $tan(\gamma)$=$\frac{c\cdot h}{a\cdot l\cdot cos(30)}$ where $a,c$ are the lattice constants. The solid red line in Fig.\ref{fig:Fig5}(d) shows the ratio of intensities calculated using the eq.\eqref{eq:ratio} with $h$ fixed to $h$=1 and plotted as a function of $l$. The calculated $l$-dependence of the ratio from $l$=0 to $l$=7 is in good agreement with the $l$-dependence of $\frac{I_{HM}}{I_{LM}}$ observed experimentally.  
\par These results confirm that the higher energy mode is caused by spin fluctuations along the c-axis while the lower energy mode is due to the fluctuations within the honeycomb plane. The excitations which fluctuate along the c-axis are gapped because the easy-plane anisotropy makes it energetically unfavourable for spins to have a component along the c-direction. The presence of a gap for the in-plane excitations suggests the existence of an additional weak easy-axis anisotropy which establishes the spin ordering direction within the $ab$-plane. Because of the 3-fold crystal symmetry of BaNi$_2$V$_2$O$_8$ there should actually be three easy-axes within the $ab$-plane which give rise to three magnetic twin domains in the ordered phase. 
\subsubsection{Single crystal magnetic excitation spectrum in the out-of-plane direction}  
\par To investigate the magnetic exchange interactions between the honeycomb layers, INS measurements were performed along the out-of-plane direction using the cold neutron spectrometers PANDA and TASP. Figure \ref{fig:Fig6}(a) shows constant wave vector scans at (1,0,$l$) over the energy range 0-5 meV  where $l$ was fixed at values from 0 to 1 in step of 0.1, thus, giving one Brillouin zone. The strong elastic signal at (1,0,$\frac{1}{2}$) corresponds to the magnetic Bragg peak while the elastic signal at (1,0,1) is due to a nuclear Bragg peak. The inelastic signals at $\approx$0.5~meV and $\approx$3.5~meV are the in-plane and out-of-plane spin-wave modes discussed in the previous section. The data clearly reveal that the upper energy mode is dispersionless within the experimental resolution. However, it is difficult to judge unambiguously whether or not the low-energy mode is dispersionless because of the strong contamination from the elastic scattering. 
\par To investigate the dispersion of the low-energy mode in more detail, a constant-wavevector scan at (1,0,$\frac{1}{2}$) was measured and compared to the constant wave-vector scan at (1,0,0). Indeed, if the low energy mode disperses along the $c$-axis, the peak positions in the constant-wavevector scans at different wave-vector transfers would be shifted with respect to each other. Figure \ref{fig:Fig5}(b) shows the constant wave-vector scan at (1,0,$\frac{1}{2}$) (open black squares) where the elastic contribution from the (1,0,$\frac{1}{2}$) magnetic Bragg peak and the flat background were subtracted. The magenta line in Fig.\ref{fig:Fig5}(b) presents the fit using the Rescal software which takes into account the instrumental resolution. The fitted peak positions are E$_1$=0.38$\pm$0.03 meV and E$_2$=3.25$\pm$0.03 meV for the energy scan at (1,0,$\frac{1}{2}$) which are the same within the error as the peak positions of E$_1$=0.41$\pm$0.03 meV and E$_2$=3.25$\pm$0.03 meV observed in the energy scan at (1,0,0). The tiny energy shift of 0.03 meV in the position of the low-energy mode can be attributed either to resolution effects or to weak modulations due to the presence of an extremely small interplane coupling. These results reveal that both modes are dispersionless along the out-of-plane direction within the experimental resolution implying that the magnetic exchange interactions between the honeycomb layers are extremely weak despite the presence of the long-range magnetic order.  
\subsection{Hamiltonian of BaNi$_2$V$_2$O$_8$} \label{subsec:SecC}  
\par In order to extract the values of the magnetic exchange interactions and to determine the anisotropy of BaNi$_2$V$_2$O$_8$, the magnetic excitation spectrum of this compound was compared to the spectrum calculated using linear spin-wave theory. The general Hamiltonian given by eq.\eqref{eq:Hamiltonian} was assumed where $J_{n}$, $J_{nn}$ and $J_{nnn}$  are the nearest,  second-nearest and  third-nearest-neighbor isotropic Heisenberg magnetic exchange interactions respectively within the honeycomb plane, and $J_{out}$ is an interplane exchange coupling (see Fig.\ref{fig:Fig1}). The interplane coupling was introduced because the magnetic structure of BaNi$_2$V$_2$O$_8$ develops long-range magnetic order below the transition temperature T$_N$=48K. The terms $D_{EP}$ and $D_{EA}$ describe the easy-plane and easy-axis single ion anisotropies of the Ni$^{2+}$ magnetic ions, respectively where the easy-axis anisotropy lies within the $ab$-plane. They are required to reproduce the energy gaps observed experimentally in the constant-wavevector scans at the dispersion minima (see Fig.\ref{fig:Fig5}). 
\begin{equation} \label{eq:Hamiltonian}
\begin{split}
H =\displaystyle\sum_{i>j} J_{n}\cdot \boldsymbol{S_i} \cdot \boldsymbol{S_j}+ \displaystyle\sum_{i>j} J_{nn}\cdot \boldsymbol{S_i} \cdot \boldsymbol{S_j} \\
+\displaystyle\sum_{i>j} J_{nnn}\cdot \boldsymbol{S_i} \cdot \boldsymbol{S_j}+ \displaystyle\sum_{i>j} J_{out}\cdot \boldsymbol{S_i} \cdot \boldsymbol{S_j} \\ 
+\displaystyle\sum_{i>j} D_{EP}\cdot S_i^{c^2}+ \displaystyle\sum_{i>j} D_{EA}\cdot S_i^{a^2}
\end{split}
\end{equation}
\par This Hamiltonian is similar to the Hamiltonians used to describe the related compounds BaNi$_2$(PO$_4$)$_2$ and BaNi$_2$(AsO$_4$)$_2$, where single ion anisotropy rather than exchange anisotropy was assumed as is appropriate for the Ni$^{2+}$ ion in an octahedral environment. The spin-wave simulations were performed for this proposed Hamiltonian using the SpinW Matlab Library \cite{SpinW}. The directions of the spins in their ground state were input into the SpinW calculation as fixed parameters based on the magnetic structure of BaNi$_2$V$_2$O$_8$ (Fig. \ref{fig:Fig1}(a)) proposed in the literature \cite{Rogado}. The initial signs of the exchange parameters $J_{n}$, $J_{n}$, $J_{nnn}$, $J_{out}$ were chosen to be consistent with this magnetic structure and their initial values were chosen to be roughly compatible with the constraint $J_{total}=J_{n}+2J_{nn}+J_{nnn} = 15\pm2$ meV extracted from the DC susceptibility measurements (eq.\eqref{eq:Jtotal}). The easy-plane anisotropy was given a positive sign ($D_{EP}>0$) to encourage the spins to lie within the honeycomb layers ($ab$-plane). The easy-axis anisotropy was given a negative sign ($D_{EA}<0$) and to encourage the spins to point along a particular direction within the honeycomb plane. Because of the three-fold rotational symmetry of the lattice of BaNi$_2$V$_2$O$_8$ about the c-axis, there are in fact three equivalent easy-axes within the honeycomb plane rotated by 120\textdegree with respect to each other. Thus below T$_N$ we expect there to be three twins with spins lying along these three different easy-axes. Each spin-wave simulation was  performed for all three twins and the results were averaged assuming that the twins have equal weights.
\par The simulations were performed iteratively varying the values of $J_{n}$, $J_{nn}$, $J_{nnn}$ ,$J_{out}$, $D_{EP}$, $D_{EA}$ about various sets of initial parameters and comparing the results to the data. The parameters $J_{n}$, $J_{nn}$, $J_{nnn}$ were found to be strongly coupled so that the range of the solutions for each of these parameters depends strongly on the values of the other parameters. To extract the whole range of possible solutions the parameters were varied manually taking into account their strong coupling. The agreement of the extracted solutions with the experimental data was checked by comparing the excitations observed in BaNi$_2$V$_2$O$_8$ (Fig. \ref{fig:Fig4}(a)-(d)) with the calculated spin-wave energies (solid red line over the data in Fig. \ref{fig:Fig4}(a)-(d)) and calculated intensities (Fig. \ref{fig:Fig4}(e)-(h)). Finally, the range of the extracted solutions was restricted to be compatible with the constraint J$_{total}$=15$\pm$2 meV. 
\par The extracted solutions reveal that the first nearest-neighbor magnetic exchange interaction $J_{n}$, is the dominant interaction with values in the range of 11.3~meV$\le$$J_{n}$$\le$13.35~meV.  The positive sign reveals that $J_{n}$ is antiferromagnetic and is compatible with the proposed magnetic structure (Fig. \ref{fig:Fig1}(a)). The second nearest-neighbor interaction $J_{nn}$ is sizeable and lies in the range 0.85 meV$\le$$J_{nn}$$\le$1.5 meV. The presence of $J_{nn}$ is necessary to reproduce the weak modulation observed experimentally at the top of the dispersion along the (h,-2h-2,0) direction (Fig. \ref{fig:Fig4}(c)). $J_{nn}$ is found to be antiferromagnetic implying that it competes with $J_{n}$ and produces a weak frustration in the magnetic system.
\par The third-neighbor interaction $J_{nnn}$, is small. It varies within the range -0.1 meV$\le J_{nnn}\le$0.4 meV and can be either FM or AFM. The presence of $J_{nnn}$ only provides a small improvement in the agreement with the experimental data and is not necessary to reproduce the observed spectrum. It was introduced to estimate the order of magnitude of other possible long-range interactions and their effect on the values of $J_{n}$ and $J_{nn}$. Indeed, the presence of an AFM $J_{nnn}$ increases the possible solutions of $J_{n}$ and $J_{nn}$ to the ranges 10.90 meV$\le$ $J_{n}$ $\le$13.35 meV and  0.85 meV$\le$ $J_{nn}$ $\le$1.65 meV respectively.   
\par The anisotropy gives rise to the energy gaps of the two spin-wave modes at the antiferromagnetic zone center (e.g. at (1,0,0) Fig.\ref{fig:Fig5}(b)), and the values of the anisotropy parameters can be extracted from the sizes of these energy gaps. The easy-plane anisotropy $D_{EP}$ must lie within the range 0.0635 meV$\le$$D_{EP}$$\le$0.0770 meV to reproduce the gap of E$_2$=3.25$\pm$0.03 meV in the higher energy mode which corresponds to out-of-plane spin fluctuations, while the easy-axis anisotropy $D_{EA}$ must be within the range -0.0011 meV$\le$$D_{EA}$$\le$-0.0009 meV to reproduce the gap of E$_1$=0.41$\pm$0.03 meV in the lower energy mode.
\par The magnetic exchange coupling between the honeycomb layers was assumed to be realized via one of three possible magnetic exchange paths $J_{out1}$, $J_{out2}$ or $J_{out3}$ which were discussed in the introduction and represented in Fig. \ref{fig:Fig1}(b) by the dashed-dotted orange, dotted green and solid cyan lines, respectively. The upper limit for each interplane coupling  was extracted by simulating the magnetic excitation spectra of BaNi$_2$V$_2$O$_8$ taking into account the conditions that (i) the upper mode is dispersionless in the out-of-plane direction within the resolution error and (ii) the energy shift of the lower mode at (1,0,0) with respect to  \((1,0,\frac{1}{2})\) does not exceed the experimentally observed value of 0.03 meV (see subsubsection \ref{sssec:SecB2}). 
\par The results reveal that $J_{out1}$ is ferromagnetic and must satisfy the condition $J_{out1}$$\ge$-0.00045 meV, while $J_{out2}$ and $J_{out3}$ are antiferromagnetic and must satisfy $J_{out2}$$\le$0.00005~meV and $J_{out3}$$\le$0.0012~meV respectively, in order to comply the conditions (i)-(ii).  Although $J_{out1}$ and $J_{out2}$ are characterized by the similar distances between Ni$^{2+}$ ions, the upper limit of $J_{out2}$ is nine times weaker than that of $J_{out1}$. However, the effective contribution of J$_{out2}$ to the interplane coupling is comparable to that of J$_{out1}$ because there are 6 J$_{out2}$ paths per Ni$^{2+}$ magnetic ion in contrast to a single J$_{out1}$. Although J$_{out3}$ can have the largest value, it  is unlikely to be responsible for the long-range magnetic order due to its large Ni$^{2+}$-Ni$^{2+}$ distance.    
\par  Figure \ref{fig:Fig3}(b) and Figure \ref{fig:Fig4}(e)-(h) show the powder and single crystal magnetic excitation spectra of BaNi$_2$V$_2$O$_8$ simulated for the Hamiltonian (eq.\eqref{eq:Hamiltonian}) with the parameters $J_{n}$=12.3~meV, $J_{nn}$=1.25~meV, $J_{nnn}$=0.2~meV, $J_{out1}$=-0.00045~meV, $D_{EP}$=0.0695~meV, $D_{EA}$=-0.0009~meV, respectively. These values lie in the middle of the ranges of possible solutions. The simulations show good agreement with the experimental data. The simulated powder spectra (Fig. \ref{fig:Fig3}(b)) extends up to 26 meV and displays the V-shaped cones of the magnon dispersions with the first minimum at $|Q|$=1.45\AA$^{-1}$$\pm$0.05\AA$^{-1}$ matching the experimental data (Fig. \ref{fig:Fig3}(a)). The simulated single crystal dispersions within the (h,k,0) scattering plane (Fig.\ref{fig:Fig4}(e)-(h)) are also consistent with the data (Fig.\ref{fig:Fig4}(a)-(d)) and the intensity modulations of the simulated modes are in good agreement with the magnetic structure factor observed experimentally. Figure \ref{fig:Fig6}(b) shows the simulated single crystal spectrum calculated along the out-of-plane (1,0,$l$) direction which again accurately matches the experimental data (Fig. \ref{fig:Fig6}(a)). 
\par The spin-wave calculations reveal that the magnetic excitation spectrum of BaNi$_2$V$_2$O$_8$ consists of the 12 modes which are shared between the two excitation branches observed in the data (Fig.\ref{fig:Fig6}(c)). Within each branch the modes are almost degenerate and are not resolved experimentally due to the finite instrumental energy resolution. The lower-energy band consists of the six spin-wave modes which extend over the energy range 0.38-0.45 meV at (1,0,$l$) (Fig.\ref{fig:Fig6}(c.2)). This leads to the finite FWHM of the lower-energy peak observed experimentally in the constant-wave vector scan at (1,0,0) (Fig.\ref{fig:Fig5}(b)). The higher-energy excitation branch also consists of the six modes but their bandwidth along (1,0,$l$) is only $\approx$0.01 meV. This explains why the higher-energy peak in the constant-wave vector scan at (1,0,0) appears resolution limited (Fig.\ref{fig:Fig5}(b)).  
\begin{table}
   \begin{tabular}{|c|c|c|c|c|c|c|}
    \hline
		\multirow{2}{*}{Compound}& \multirow{2}{*}{$J_{eff}$,($\frac{J_{eff}}{k_B}$)}& \multirow{2}{*}{$\frac{J_{out}}{J_{eff}}$} & \multirow{2}{*}{$\frac{T_N}{J_{eff}}$} &\multirow{2}{*}{$\frac{D_{EP}}{J_{eff}}$} & \multirow{2}{*}{$\frac{T_{ani}}{J_{eff}}$} & \multirow{2}{*}{$\frac{T_{XY}}{J_{eff}}$}\\
		\hhline{~~~~~~}            &  &  & & & \\
		\hline 
    \multirow{2}{*}{BaNi$_2$V$_2$O$_8$} & 10meV &\multirow{2}{*}{0.000045}&\multirow{2}{*}{0.4}& \multirow{2}{*}{0.007} & \multirow{2}{*}{0.7} & \multirow{2}{*}{0.45} \\
    \hhline{~~~~~~}            & ($\approx$116K) &  & & & & \\
		\hline 
    \multirow{2}{*}{BaNi$_2$P$_2$O$_8$} & 1.792meV &\multirow{2}{*}{0.00009}&\multirow{2}{*}{1.18} & \multirow{2}{*}{0.337} & \multirow{2}{*}{1.5} & \multirow{2}{*}{---} \\
    \hhline{~~~~~~}            & ($\approx$20.8K) &  & & & & \\
				\hline 
    \multirow{2}{*}{BaNi$_2$As$_2$O$_8$} & 1.896meV & \multirow{2}{*}{---} & \multirow{2}{*}{0.89} & \multirow{2}{*}{0.318} & \multirow{2}{*}{1.45} & \multirow{2}{*}{---} \\
    \hhline{~~~~~~}            & ($\approx$22K)  &  & & & & \\
		\hline 
  \end{tabular} 
	\caption{The effective in-plane magnetic exchange interaction $J_{eff}$ calculated for BaNi$_2$M$_2$O$_8$ (M=V,P,As) assuming that each bond is counted only once. The values of the interplane coupling $J_{out1}$, ordering temperature T$_N$, easy-plane single-ion anisotropy D$_{EP}$ (averaged value), anisotropy temperature T$_{ani}$, and planar regime temperature T$_{XY}$ are also given after scaling by $J_{eff}$.}
	\label{tab:Hamiltonian}
\end{table} 
\section{Discussion}
\par The observed magnetic properties of BaNi$_2$V$_2$O$_8$ can be compared to the magnetic properties of the related nickel-based S=1 honeycomb antiferromagnets BaNi$_2$(PO$_4$)$_2$ and BaNi$_2$(AsO$_4$)$_2$, which also have easy-plane anisotropy. Of particular interest is the comparison of their anisotropies and interlayer couplings, and also their N\'{e}el and anisotropy temperatures in order to determine the existence and extent of their various magnetic regimes. Because these compounds have different magnetic structures and exchange interactions (e.g. in BaNi$_2$(PO$_4$)$_2$ the strongest interaction is between third nearest neighbor\cite{Jongh}) it is necessary to find a way to compare them that takes these differences into account.
\par For each compound an effective first neighbor magnetic exchange interaction  $J_{eff}$ is defined, which represents the three nearest-neighbor exchange interactions and is compatible with the magnetic structure. It is given by:  
\begin{equation}\label{eq:Jeff}
{J_{eff}=\frac{\sum_{i}^{3} (z^{+}_i|J_i|-z^{-}_i|J_i|)}{N_{nearest}}}
\end{equation}
where $|J_{i}|$ is the absolute value of the exchange interaction between a magnetic ion and its $i$$^{th}$-nearest neighbor, $N_{nearest}$ is the number of first neighbor magnetic ions which is $N_{nearest}$=3 for a honeycomb lattice. $z^+_i$ and $z^-_i$ are the numbers of bonds between a magnetic ion and its $i^{\it{th}}$-nearest neighbors which reinforce and oppose the magnetic structure respectively. In the case of BaNi$_2$V$_2$O$_8$, the first and third neigbor interactions $J_n$ and $J_{nnn}$ which are both AFM, reinforce the magnetic structure, because as shown in Fig.\ref{fig:Fig1}(a), first and third neighbor spins are aligned antiparallel at low temperatures ($z^{+}_{n}=z^{+}_{nnn}=3$ and $z^{-}_{n}=z^{-}_{nnn}=0$). On the other hand the second neighbor interaction $J_{nn}$ which is also AFM, opposes the structure because second neighbor spins are parallel ($z^{+}_{nn}=0$ and $z^{-}_{nn}=6$) .
\par Table \ref{tab:Hamiltonian} summarizes the values of $J_{eff}$ for BaNi$_2$M$_2$O$_8$ (M=V,P,As). The value for BaNi$_2$V$_2$O$_8$ is found to be  $J_{eff}$$\approx$10meV and was calculated using the values of $J_n$=12.3meV, $J_{nn}$=1.25meV and $J_{nnn}$=0.2meV extracted from fitting the magnetic excitation spectrum. The values of $J_{eff}$ = 1.792meV and $J_{eff}$ = 1.896meV for BaNi$_2$P$_2$O$_8$ and BaNi$_2$As$_2$O$_8$ respectively, were estimated using the corresponding magnetic structures and Hamiltonians reported in the literature \cite{Jongh,Regnaulall}.  
\par The value of $J_{eff}$ for BaNi$_2$V$_2$O$_8$ is much greater than the values for BaNi$_2$P$_2$O$_8$ and BaNi$_2$As$_2$O$_8$, therefore to compare the important temperatures and interaction energies of BaNi$_2$V$_2$O$_8$ to those of the other compounds, these quantities are first scaled by $J_{eff}$ as listed in Table \ref{tab:Hamiltonian}. The scaled interplane coupling in  BaNi$_2$V$_2$O$_8$ ($J_{out} = 0.000045J_{eff}$) is weaker than in BaNi$_2$P$_2$O$_8$ ($J_{out} = 0.00009J_{eff}$). As a consequence the ordering temperature T$_{N} = 0.4J_{eff}$/k$_B$ in BaNi$_2$V$_2$O$_8$, is also suppressed in the relative scale with respect to T$_{N}~=1.18J_{eff}$/k$_B$ and T$_{N}~=~0.89J_{eff}$/k$_B$ observed in BaNi$_2$P$_2$O$_8$ and BaNi$_2$As$_2$O$_8$ respectively. The scaled easy-plane anisotropy in BaNi$_2$V$_2$O$_8$ $D_{EP}\approx 0.007J_{eff}$, is much weaker than the values $D_{EP}{\approx 0.337}J_{eff}$ and $D_{EP}{\approx0.318}J_{eff}$ estimated for BaNi$_2$P$_2$O$_8$ and  BaNi$_2$As$_2$O$_8$ respectively. This leads to the appearance of anisotropic magnetic behaviour in BaNi$_2$V$_2$O$_8$ at lower relative temperature than in the other two compounds. Indeed, anisotropic magnetism in BaNi$_2$V$_2$O$_8$ appears in the susceptibility at T$_{ani}$=80K below which $\chi_{H\perp c}$ and $\chi_{H||c}$ split, this corresponds to T$_{ani}$$\approx$0.7$J_{eff}$/k$_B$ in the relative temperature scale. For comparison, in both BaNi$_2$P$_2$O$_8$ and  BaNi$_2$As$_2$O$_8$ the $\chi_{H\perp c}$ and $\chi_{H||c}$ start to split at T$_{ani}$=32K \cite{Jongh} which in the relative temperature scale correspond to T$_{ani}$$\approx$3J$_{eff}$/k$_B$ and T$_{ani}$$\approx$2.9J$_{eff}$/k$_B$, respectively. 
\begin{figure}
\includegraphics[width=1\linewidth]{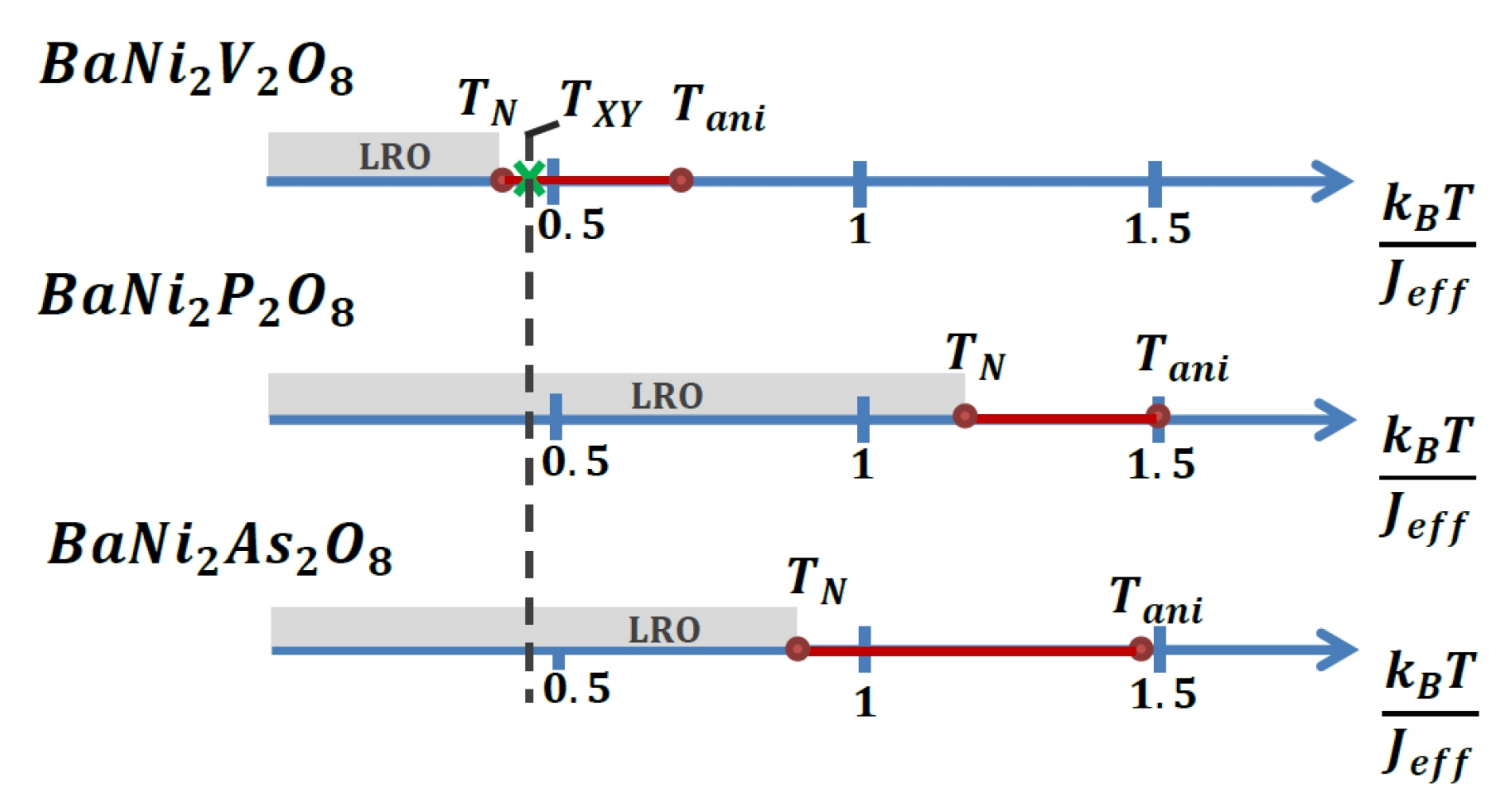} 
\caption{The magnetic regimes of BaNi$_2$V$_2$O$_8$, BaNi$_2$P$_2$O$_8$ and BaNi$_2$As$_2$O$_8$ as a function of the relative temperature scale (k$_B$T/$J_{eff}$). The N\'{e}el temperature T$_{N}$ and the temperature T$_{ani}$ where the magnetic anisotropy first appears are indicated by the red filled circles. The gray area below T$_{N}$ indicats the long-range magnetically ordered phase. The temperature T$_{XY}$, at which $\chi_{H||c}$ in BaNi$_2$V$_2$O$_8$ has a minimum associated with EP anisotropy regime, is marked by the green cross. The black-dashed line indicates the position of T$_{XY}$ with respect to BaNi$_2$P$_2$O$_8$ and BaNi$_2$As$_2$O$_8$.} 
\label{fig:Fig7} 
\end{figure}
\par The experimental susceptibility of BaNi$_2$V$_2$O$_8$ can be compared with the results of Quantum Monte Carlo (QMC) simulations performed by Cuccoli et. al.\cite{Cuccoli,Cuccoli2} for the antiferromagnetic spin-1/2 2D square lattice with weakly Ising  (XXZ-type) interactions. Although the simulations do not include the interplane coupling or the easy-axis anisotropy the qualitative agreement with $\chi_{H\perp c}$ and $\chi_{H||c}$ of BaNi$_2$V$_2$O$_8$ is remarkable (see fig. 1 of Ref \onlinecite{Cuccoli2}). In particular, the QMC simulations predict that the in-plane and out-of-plane susceptibilities are isotropic at high temperatures but split from each other as temperature decreases revealing anisotropic behavior. This is, indeed, what was observed in the experimental susceptibility of BaNi$_2$V$_2$O$_8$ below T$_{ani}$=80K (T$_{ani}$$\approx$ 0.7$J_{eff}$/k$_B$). Moreover, according to QMC simulations, the system displays a crossover to a regime where the easy-plane anisotropy dominates. This crossover manifests itself as a minimum at a finite temperature T$_{co}$=0.3$J_n$/k$_{B}$ below which the out-of-plane susceptibility reduces only a little and then becomes approximately constant while the in-plane susceptibility drops steadily \cite{Cuccoli,Cuccoli2}. This is in a good agreement with $\chi_{H||c}$ in BaNi$_2$V$_2$O$_8$ which reveals a minimum at T$_{XY}$=0.45$J_{eff}$/k$_B$ in the relative scale (T$_{XY}$=52K in absolute scale) and then increases slightly and becomes constant with further temperature decrease, in contrast to $\chi_{H\perp c}$, which decreases continuously. The minimum observed in the $\chi_{H||c}$ of BaNi$_2$V$_2$O$_8$ at T$_{XY}$=0.45$J_{eff}$/k$_B$(Fig.\ref{fig:Fig2}) can be assigned to the minimum at T$_{co}$ predicted by QMC \cite{Cuccoli,Cuccoli2} implying a crossover to a regime where BaNi$_2$V$_2$O$_8$ displays strong 2D easy-plane anisotropy.
\par It is remarkable that this characteristic minimum associated with a crossover to an easy-plane dominated regime is experimentally observed only in the $\chi_{H||c}$ of BaNi$_2$V$_2$O$_8$ while susceptibilities of BaNi$_2$P$_2$O$_8$ and BaNi$_2$As$_2$O$_8$ reveal only the splitting of $\chi_{H\perp c}$ and $\chi_{H||c}$ below T$_{ani}$. It can be understood by comparing the value of T$_N$ in BaNi$_2$V$_2$O$_8$, BaNi$_2$P$_2$O$_8$ and BaNi$_2$As$_2$O$_8$ in a relative temperature scale. Indeed, for all compounds, quasi-2D easy-plane anisotropy should be observable over the temperature range T$_N$-T$_{ani}$, while below T$_N$ 3D magnetic behavior dominates and above T$_{ani}$ the effect of the easy-palne anisotropy is overcome by the thermal energy. The value of T$_{XY}$ for BaNi$_2$P$_2$O$_8$ and BaNi$_2$As$_2$O$_8$ can be expected to be similar to the value  T$_{XY}$=0.45$J_{eff}$/k$_B$ extracted for BaNi$_2$V$_2$O$_8$. Because in both BaNi$_2$P$_2$O$_8$ and BaNi$_2$As$_2$O$_8$ T$_N~\approx~1 Jeff/k_B$ lies at higher relative temperatures than T$_{XY}$=0.45$J_{eff}$/k$_B$, the easy-plane regime in these compounds is overcome by the long-range ordered phase and the predicted minimum in $\chi_{H||c}$ is not present in their susceptibility data\cite{Jongh}. 

\section{Conclusions}
\par We have performed a comprehensive investigation of the novel spin-1, honeycomb, antiferromagnet BaNi$_2$V$_2$O$_8$ employing inelastic neutron scattering and DC susceptibility measurements. The observed low-temperature susceptibility data $\chi_{H\perp c}$ and $\chi_{H||c}$, are in good agreement with the experimental results published previously\cite{Rogado} and reveal that BaNi$_2$V$_2$O$_8$ behaves as a quasi-2D Heisenberg antiferromagnet with easy-plane anisotropy. 
\par The transition to long-range antiferromagnetic order is evident only as a subtle inflection point detected in $\chi_{H||c}$ at T$_N$=48K$\pm$0.5K in agreement with the weak features reported previously at the same temperature in both $\chi_{H||c}$\cite{Rogado} and heat capacity data\cite{Knafo}. It is also consistent with the phase transition to the antiferromagnetic ordered state observed at T$_N$=47.5K-47.75K by neutron diffraction measurements on the same crystals \cite{FiniteTBaNiVO}. 
\par The crossover to the regime dominated by easy-plane anisotropy is observed in BaNi$_2$V$_2$O$_8$ at T$_{XY}$=52K and is characterized by a minimum in $\chi_{H||c}$ (see Fig.\ref{fig:Fig7}) which is in agreement with the predictions of Quantum Monte Carlo Simulations performed for the S=$\frac{1}{2}$ 2D XXZ square lattice\cite{Cuccoli}. The corresponding relative temperature T$_{XY}$=0.45$J_{eff}$/k$_B$, is also found to be in general agreement with T$_{XY}^{QMC} \approx 0.3 J_n$/k$_B$ predicted by QMC.
\par  The analysis of the high temperature susceptibility data allows the Curie-Weiss temperature ~$\Theta_{CW}=-542~\pm$~76~K and Curie constant to be extracted for the first time. The effective magnetic moment of the Ni$^{2+}$ion was estimated from the Curie constant and reveals that the orbital angular momentum of the magnetic Ni$^{2+}$ ion is almost but not completely quenched. The fact that the orbital angular momentum is partially recovered indicates the presence of a weak spin-orbit coupling and is consistent with the single-ion anisotropy observed in this magnetic system \cite{Yosida}.       
\par The INS measurements reveal that BaNi$_2$V$_2$O$_8$ has magnetic spin-wave excitations which disperse within the honeycomb plane and are dispersionless (within resolution error) in the out-of-plane direction confirming that BaNi$_2$V$_2$O$_8$ is an almost ideal 2D magnetic system. The observed magnetic excitation spectrum of BaNi$_2$V$_2$O$_8$ consists of two gapped spin-wave branches due to in-plane and out-of-plane spin-wave fluctuations. The analysis of the relative intensities of these branches as a function of wavevector transfer reveals that the lower-energy branch is due to fluctuations within the honeycomb plane while the higher-energy branch is due to fluctuations along the c-axis. The fact that the out-of-plane fluctuations have a significant energy gap confirms the presence of a dominant easy-plane anisotropy which makes these fluctuations energetically more costly to excite. The smaller energy gap of the in-plane spin fluctuations implies the presence of a weak easy-axis anisotropy within the honeycomb plane. The extracted values of the energy gaps E$_1$=0.41$\pm$0.03meV and E$_2$=3.25$\pm$0.03meV, are in qualitative agreement with the values of E$_1$=0.2meV and E$_2$=3meV briefly mentioned in the paper by Knafo et. al. \onlinecite{Knafo}. However, since no data is shown, it is difficult to make a quantitative comparison with our results. 
\par Both the powder and single crystal magnetic excitation spectra of BaNi$_2$V$_2$O$_8$ were reproduced well by linear spin-wave theory allowing the Hamiltonian of BaNi$_2$V$_2$O$_8$ to be determined for the first time. The solution reveals that the dominant term is the nearest-neighbor in-plane exchange interaction $J_{n}$, which is AFM and lies within the range of 10.9meV$\le$$J_{n}$$\le$13.35meV. This value of $J_{n}$ is considerably larger than the value $J_1$$\approx$8.3meV published by Rogado et. al. \onlinecite{Rogado} which was extracted from DC susceptibility using the High Temperature Series Expansion approach. The discrepancy can be attributed to the fact that the second neighbor interaction was not considered and the data used for the comparison were collected at relatively low temperatures (T$\le$300K).
\par The second $J_{nn}$ and third  $J_{nnn}$ neighbor in-plane interactions were found to be much smaller and to vary within the ranges of 0.85meV$\le$$J_{nn}$$\le$1.65meV and -0.1meV$\le$$J_{nnn}$$\le$0.4meV, respectively. $J_{nn}$ is antiferromagnetic and opposes the magnetic structure of BaNi$_2$V$_2$O$_8$ introducing a weak frustration to the system. The third neighbour interaction $J_{nnn}$ is not necessary to reproduce the experimental data and represents the size of the contributions from distant neighbors.  
\par The values of the easy-plane and easy-axis anisotropies are extracted to be within the ranges 0.0635$\le$$D_{EP}$$\le$0.0770~meV and -0.0011$\le$$D_{EA}$$\le$-0.0009meV, respectively. The dominance of the easy-plane anisotropy over the easy-axis anisotropy confirms the planar character of the magnetism in this compound. Although BaNi$_2$V$_2$O$_8$ displays long-range magnetic order, the interplane coupling $J_{out}$ was found to be surprisingly small and to vary within the energy range -0.00045meV $\le$ J$_{out}$$\le$0.0012meV where the sign of the $J_{out}$ depends on the exchange path responsible for this coupling. The very weak $J_{out}$ explains the absence of clear signatures of the transition to long-range magnetic order in the DC susceptibility and heat capacity\cite{Knafo} data. 
\par The presence of the long-range magnetic order despite such a weak interplane coupling can be attributed to the presence of weak easy-axis anisotropy. Indeed, Knafo et. al.\cite{Knafo} demonstrate the importance of the easy-axis anisotropy for inducing the long-range magnetically ordered state in BaNi$_2$V$_2$O$_8$ by exploring the effects of the field- and pressure-induced anisotropies on the ordering temperature T$_N$. Thus, a combination of a very weak interplane coupling and a very weak easy-axis anisotropy may together be sufficient to induce long-range magnetic order at the finite temperature of T$_N$=48$\pm$0.5K in this planar antiferromagnet.
\par To summmarize, BaNi$_2$V$_2$O$_8$ is a highly 2D antiferromagnet with dominant easy-plane anisotropy and is, thus, a promising candidate for the further investigation of BKT behavior\cite{Heinrich,Waibel2015}. The temperature region T$_N$-T$_{XY}$, where the easy-plane anisotropy dominates but which is above the transition to 3D order, is especially interesting for the observation of vortex excitations and BKT phenomena.
\begin{acknowledgments}
\emph{Acknowledgments} - The susceptibility measurement were perform at the Core Lab for Quantum Materials, HZB; We thank K. Siemensmeyer for helping with these measurements. We also thank S. Toth for his help with the SpinW simulations. M. M. was partly supported by Marie Sklodowska Curie Action, International Career Grant through the European Commission and Swedish Research Council (VR), Grant No. INCA-2014-6426 as well as a VR neutron project grant (BIFROST, Dnr. 2016-06955).
\end{acknowledgments}
\appendix
\section{}
\par In this appendix the ratio of the neutron scattering intensities of the two spin-wave modes at (1,0,L) in BaNi$_2$V$_2$O$_8$ is calculated as a function of wavevector L. This calculation takes into account the known magnetic structure of this compound and is averaged of the three magnetic twin domains. The results are used in section \ref{sssec:SecB2} to verify the fluctuation directions of the two spin-wave modes and provide quantitative agreement with the data.  
\begin{figure} 
\includegraphics[width=0.8\linewidth]{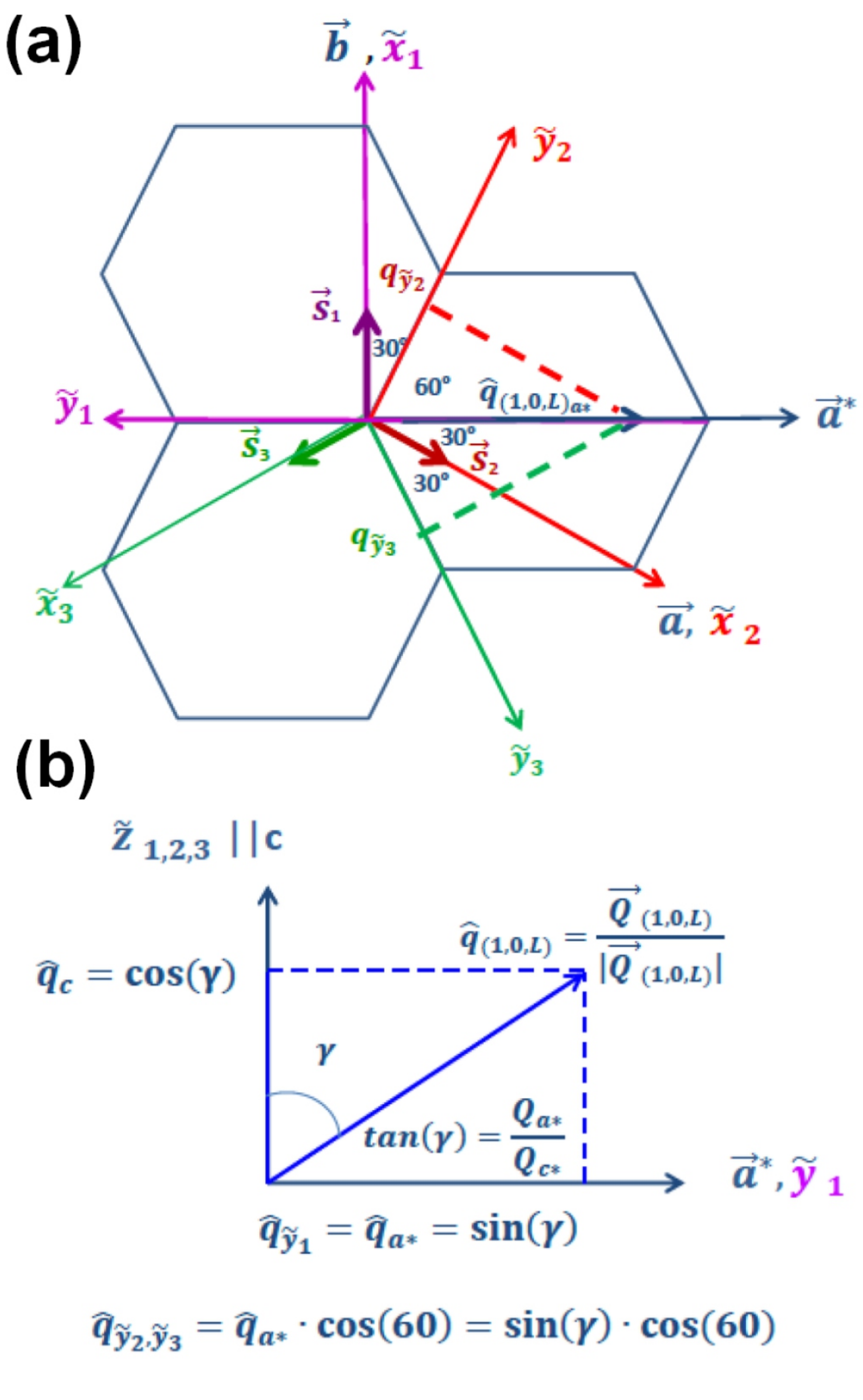} 
\caption{(a) Magnetic structure of BaNi$_2$V$_2$O$_8$ in the honeycomb plane. The spin (thick arrows) and coordinate systems (thin arrows) for the three twinned domains are indicated by the magenta, red and green colors respectively. The projection of the (1,0,$l$) wavevector transfer onto the spin direction of each twin is also shown.  b) A sketch of the $a^{\star}c^{\star}$-plane and wavevector transfer $\vec{Q}$.}
\label{fig:FigAppendix}
\end{figure}
\par Figure \ref{fig:FigAppendix}(a) shows the direction of the spins in BaNi$_2$V$_2$O$_8$within the honeycomb ($ab$) plane below T$_N$. The thick magenta, red and green arrows labelled $\vec{s_1}$, $\vec{s_2}$ and $\vec{s_3}$ respectively, indicate the spin directions of the three twins. The local coordinate systems $\tilde{x}_n,\tilde{y}_n,\tilde{z}_n$ (n=1,2,3) for all three twins are also plotted using the same color coding. Each coordinate system is constructed so that $\tilde{x}_n ||\vec{s_n}$, $\tilde{y}_n \perp\vec{s_n}$, $\tilde{z}$=$\tilde{x}\times\tilde{y}$ and $\tilde{z}_n || c$-axis. 
\par For each twin the neutron scattering cross section is proportional to the relation \cite{Squires}:
\begin{subequations} \label{eq:Ratio1}
\begin{align}
I &\propto (1-\hat{q}^2_{\tilde{x}})\cdot S^{\tilde{x}\tilde{x}}(Q,w)  \\
  &+ (1-\hat{q}^2_{\tilde{y}})\cdot S^{\tilde{y}\tilde{y}}(Q,w)\\
  &+ (1-\hat{q}^2_{\tilde{z}})\cdot S^{\tilde{z}\tilde{z}}(Q,w)
\end{align}
\end{subequations}
\par Here, $\hat{q}$ is a unit vector in the direction of $\vec{Q}$ and $\hat{q}_{\tilde{x}}$, $\hat{q}_{\tilde{y}}$, $\hat{q}_{\tilde{z}}$ are its components within the local coordinate system $\tilde{x}$,$\tilde{y}$,$\tilde{z}$ of that twin. In the ordered phase the first term in equation \eqref{eq:Ratio1} is responsible for the elastic magnetic signal, while the second and third terms give the intensity of the inelastic neutron scattering due to spin-fluctuations perpendicular to the order along the $\tilde{y}$ and $\tilde{z}$ directions. Therefore the ratio of the intensities $I_{\tilde{z}}$ and $I_{\tilde{y}}$, of the modes due to the spin fluctuations along the $\tilde{y}$ and $\tilde{z}$ directions respectively, is proportional to
\begin{equation} \label{eq:Ratio2}
{\frac{I_{\tilde{z}}}{I_{\tilde{y}}} \propto \frac{(1-\hat{q}^2_{\tilde{z}})}{(1-\hat{q}^2_{\tilde{y}})}}
\end{equation}
\par If there are domains, the terms $\hat{q}^2_{\tilde{z}}$ and $\hat{q}^2_{\tilde{y}}$ in eq.\eqref{eq:Ratio1} should be replaced by their averaged values over the domains $\langle\hat{q}^2_{\tilde{z}}\rangle$ and $\langle\hat{q}^2_{\tilde{y}}\rangle$. 
\par In BaNi$_2$V$_2$O$_8$,  $\tilde{z}_n$ is parallel to the c-axis for all twins so that $\hat{q}_{\tilde{z}_1}$=$\hat{q}_{\tilde{z}_2}$=$\hat{q}_{\tilde{z}_3}$=$\hat{q}_{c}$ where $\hat{q}_c$ is the projection of the unit wavevector $\hat{q}$ on the c-axis. 
Thus, the ratio of the scattering intensities $I_{c}$ and $I_{(a,b)}$, due to spin-fluctuations along the $c$-axis and within the honeycomb plane respectively, is proportional to:
\begin{equation} \label{eq:Ratio2a}
{\frac{I_{c}}{I_{(a,b)}} \propto =\frac{(1-\hat{q}^2_{c})}{(1-\langle\hat{q}^2_{\tilde{y}}\rangle)}}
\end{equation}
\par The projection $\hat{q}_{c}$ of the unit wavevector $\hat{q}$ on the c-axis and its averaged value over the twins are given by 
\begin{equation} \label{eq:Ratio3}
\begin{split}
\hat{q}_c &= cos(\gamma) \\
\langle\hat{q}^2_{c}\rangle & =cos^2(\gamma)
\end{split} 
\end{equation}
\par where, $\gamma$ is the angle between the wavector transfer $\vec{Q}$ and the c-axis which is $\gamma=atan(\frac{Q_{a^*}}{Q_{c^*}})$ where $Q_{a^*}$=$\frac{2\pi\cdot h}{a\cdot cos(30)}$ and $Q_{c^*}$=$\frac{2\pi\cdot l}{c}$ (see Fig. \ref{fig:FigAppendix}(b)). 
\par The projection of $\hat{q}$ on the ${\tilde{y}_n}$-axis can be found for each twin using the relation: 
\begin{equation}
{\hat{q}_{\tilde{y}_n} =\hat{q}_{a^*}\cdot cos(\beta_n)=sin(\gamma)\cdot cos(\beta_n)}
\end{equation}
 where, $\beta_n$ is the angle between the $a^*$-axis and the $\tilde{y}_n$-axis for the the n$^{th}$ twin. As shown in Fig. \ref{fig:FigAppendix}(a) $\beta_1$=180$\degree$, $\beta_2$=60$\degree$ and $\beta_3$=-60$\degree$ for the first, second and third twins, respectively. Therefore assuming that the three twins are equally populated: 
\begin{equation} \label{eq:Ratio4}
\begin{split}
\hat{q}_{\tilde{y}_1} &= -sin(\gamma) \\
\hat{q}_{\tilde{y}_2}  &= \hat{q}_{\tilde{y}_3}= sin(\gamma)\cdot cos(60) \\
\langle\hat{q}^2_{\tilde{y}}\rangle &= \frac{1}{3}\cdot sin^2(\gamma)+ \frac{2}{3}\cdot sin^2(\gamma)\cdot cos^2(60)
\end{split}
\end{equation}
 
The ratio of the intensity of the mode due to the out-of-plane spin fluctuations ($I_{c}$) to the intensity of the mode due to in-plane spin fluctuations ($I_{(a,b)}$) averaged over all twins can be obtained by the substitution of the equations eq.\eqref{eq:Ratio4} and eq.\eqref{eq:Ratio3} into eq.\eqref{eq:Ratio2a} giving:

\begin{equation}
{\frac{I_{c}}{I_{(a,b)}} \propto \frac{(1-cos^2(\gamma))}{(1-(\frac{1}{3}\cdot sin^2(\gamma)+ \frac{2}{3}\cdot sin^2(\gamma)\cdot cos^2(60)))}}
\end{equation}  
 
%

%\bibliography{referencesBaNiVO2}

\end{document}